\documentclass[12pt]{article}
\usepackage{graphicx}
\def\laq{\raise 0.4ex\hbox{$<$}\kern -0.8em\lower 0.62
ex\hbox{$\sim$}}
\def\gaq{\raise 0.4ex\hbox{$>$}\kern -0.7em\lower 0.62
ex\hbox{$\sim$}}

\begin{document}
 
\begin{titlepage}
\begin{flushright}
CERN-PH-TH/2008-232
\end{flushright}
\vspace*{1cm}
 
\begin{center}
 
{\large{\bf Cosmic polarimetry in magnetoactive plasmas}}
\vskip1.cm
 
Massimo Giovannini$^{a,b}$\footnote{Electronic address: massimo.giovannini@cern.ch} and Kerstin E. Kunze$^{a,c}$
\footnote{Electronic address: kkunze@usal.es}
 
\vskip1.5cm
{\sl $^a$  Department of Physics, Theory Division, CERN, 1211 Geneva 23, Switzerland}
\vskip 0.2cm
{\sl $^b$ INFN, Section of Milan-Bicocca, 20126 Milan, Italy}
\vskip 0.2cm
{\sl $^c$  Departamento de F\'\i sica Fundamental, \\
 Universidad de Salamanca, Plaza de la Merced s/n, E-37008 Salamanca, Spain}

\vspace*{1cm}

\begin{abstract}
Polarimetry of the Cosmic Microwave Background (CMB) 
represents one of the possible diagnostics aimed at  testing large-scale 
magnetism at the epoch of the photon decoupling. 
The propagation of electromagnetic disturbances in a magnetized plasma leads naturally to a B-mode polarization  whose angular power spectrum is hereby computed both analytically and numerically. Combined analyses of all the publicly available data on the B-mode polarization are presented, for the first time, in the light of the magnetized $\Lambda$CDM scenario. Novel constraints on pre-equality magnetism are also derived in view of the current and expected sensitivities to the B-mode polarization.
\end{abstract}
 
\end{center}
\end{titlepage}
\newpage
\renewcommand{\theequation}{1.\arabic{equation}}
\setcounter{equation}{0}
\section{Formulation of the problem}
\label{sec1}
When (linearly) polarized electromagnetic radiation propagates in a diamagnetic material, the polarization plane of the wave changes by an 
angle $\Delta\Phi(\lambda, B_{\parallel})$
\begin{equation}
\Delta\,\Phi(\lambda, B_{\parallel}) = {\mathcal V}(\lambda,T...)\,\, B_{\parallel}\, L,
\label{verdet}
\end{equation}
where ${\mathcal V}$ denotes the Verdet constant,
 $B_{\parallel}$ characterizes the magnetic field intensity along the direction 
 of propagation of the electromagnetic radiation of wavelength 
 $\lambda$; $L$ is the distance travelled by the wave within the medium. 
 The Verdet constant ${\mathcal V}$ not  only depends
  upon the wavelength but also on the specifics 
 of the given material such as, for instance,  the temperature and the charge 
 concentration.   In electromagnetic plasmas, 
  the kinetic energy of the charge carriers 
   dominates against the potential energy,  and 
   the  Verdet constant can be computed in simple terms:
\begin{equation}
\Delta \Phi = \mathrm{RM}(\tilde{n}_{\mathrm{e}}, B_{\parallel})\,
 \lambda^2 , \qquad \mathrm{RM} = \frac{e^3}{2\pi\, m_{\mathrm{e}}^2 }\int_{0}^{L} \tilde{n}_{\mathrm{e}}
\,\, B_{\parallel} \, ds, 
\label{plasma}
\end{equation}
where $m_{\mathrm{e}}$ is the electron mass, $\tilde{n}_{\mathrm{e}}$ is the charge concentration of the electrons and the integral allows 
for the spatial variation of the magnetic field intensity as well as of the charge concentration. Equation (\ref{plasma}) holds under two 
physical assumptions (see, e.g. \cite{alfven,krall}):
\begin{itemize}
\item{} the plasma is globally neutral, i.e.  
 the charge concentrations of the electrons and of the ions 
are balanced, i.e. $\tilde{n}_{\mathrm{e}} = \tilde{n}_{\mathrm{i}} = \tilde{n}_{0}$  (where $\tilde{n}_{0}$ denotes generically the common value of $\tilde{n}_{\mathrm{e}}$ and $\tilde{n}_{\mathrm{i}}$);
\item{} the plasma is cold, i.e. the kinetic temperature of the electrons 
and of the ions is always much smaller than the corresponding masses.
\end{itemize}
Since the  plasma is cold (i.e. non-relativistic) and globally neutral 
the values of the plasma and Larmor frequencies for the electrons are always smaller than the corresponding quantities but computed for the ions: this is why, in Eq. (\ref{plasma}) only the electron mass and the electron concentration appears. In the case of a  tokamak Eq. (\ref{plasma}) 
is often used to determine the charge 
concentration $\tilde{n}_{\mathrm{e}}$.  
Linearly polarized electromagnetic radiation can be sent through the tokamak and the $\Delta\Phi(\lambda^2)$ can be directly measured, for instance, as a function of the frequency 
of the incident radiation. The latter experimental information is sufficient  to determine, at least approximately, the 
electron concentration $\tilde{n}_{\mathrm{e}}$ since the geometry and intensity of the magnetic field are known from the design of the instrument.  

Given the profile of the electron concentration it is 
relevant, in diverse circumstances, to assess the magnetic field 
intensity.  This is, in a nutshell, one of the objectives of cosmic polarimetry: the polarization properties of the cosmic microwave background radiation (CMB in what follows) are exploited as a diagnostic for the existence of large-scale magnetic fields. 
The polarization observable of the radiation field are 
customarily parametrized  in terms of  the Stokes parameters 
\cite{jackson} whose evolution  can be written in a schematic form which is closely analog to the set of equations often employed to describe the propagation of polarized radiation through stellar atmospheres 
\cite{st1,st2,st3}:  
\begin{eqnarray}
&& \frac{d I}{d\tau} + \epsilon' I = \epsilon' {\mathcal S}_{I}, 
\label{heat1}\\
&& \frac{d Q}{d\tau} + \epsilon' Q = \epsilon' {\mathcal S}_{Q}  + 2 \epsilon' \kappa_{\mathrm{F}} \, U,
\label{heat2}\\
&& \frac{d U}{ d\tau} + \epsilon' U = - 2 \epsilon' \kappa_{\mathrm{F}} \, Q,
\label{heat3}
\end{eqnarray}
where
\begin{itemize}
\item{} $\epsilon'$ is the differential optical depth of the system (mainly due, in the CMB case, to photon-electron 
scattering) and $\kappa_{\mathrm{F}}$ is the Faraday rotation rate, i.e. $\kappa_{\mathrm{F}} = 3\,\lambda^2\, ( \hat{n} \cdot \vec{B})/(16\,\pi^2 \, e)$;
\item{} ${\mathcal S}_{I}$ and ${\mathcal S}_{Q}$ are the source terms for the intensity 
and for the polarization;
\item{} within the present notations, the (rotationally invariant) $I$ and $V$ Stokes parameters describe, respectively, the intensity of the radiation field and the circular polarization; the $Q$ and $U$ Stokes parameters are not rotationally invariant and are both sensitive to the linear polarization.
\end{itemize}
Furthermore, according to the standard conventions, we also have:
\begin{equation}
\Phi = \frac{1}{2} \arctan{\biggl(\frac{U}{Q}\biggr)}, \qquad P= \sqrt{ Q^2 + U^2}.
\label{heat4}
\end{equation}
Equations (\ref{heat1})--(\ref{heat3}) are the flat-space 
 analog of the evolution equations studied in the present 
 paper. In Eqs. (\ref{heat1}), (\ref{heat2}) and (\ref{heat3}) the convective 
derivative with respect to $\tau$ is actually a total derivative with respect to the cosmic time coordinate which receives contribution also from the fluctuations of the geometry.

The large-scale fluctuations of the geometry induce 
computable inhomogeneities on the Stokes parameters and they will be denoted, respectively, as\footnote{In the spherical basis $\hat{n} = (\vartheta,\, \varphi)$ and $d\hat{n} = \sin{\vartheta} \,d\vartheta\, d\varphi = 
- d\cos{\vartheta} d\varphi$. The notation
$\mu = \cos{\vartheta}$ will be adopted.} $\Delta_{\mathrm{I}}(\hat{n},\tau)$, $\Delta_{\mathrm{Q}}(\hat{n},\tau)$ 
and $\Delta_{\mathrm{U}}(\hat{n},\tau)$. The experimental observations on the properties of the CMB polarization are expressed not in terms 
of $\Delta_{\mathrm{Q}}(\hat{n},\tau)$ and $\Delta_{\mathrm{U}}(\hat{n},\tau)$ but rather in terms of the corresponding E-mode and B-mode polarizations. The rationale for this common practice stems directly from the observation that the brightness perturbations of the Stokes parameters sensitive to linear polarization 
(i.e. $\Delta_{\mathrm{Q}}(\hat{n},\tau)$ and $\Delta_{\mathrm{U}}(\hat{n},\tau)$) are not invariant under rotations on a plane orthogonal $\hat{n}$. Consequently, their orthogonal combinations 
\begin{equation}
\Delta_{\pm}(\hat{n},\tau) = \Delta_{\mathrm{Q}}(\hat{n},\tau) \pm i \Delta_{\mathrm{U}}(\hat{n},\tau),
\label{int1}
\end{equation}
transform, respectively, as fluctuations of spin-weight $\pm 2$ \cite{zalda} (see also \cite{sud}). Owing to this observation,  $\Delta_{\pm}(\hat{n},\tau)$ can be expanded in terms of spin-$\pm2$ spherical harmonics 
$_{\pm 2}Y_{\ell\,m}(\hat{n})$, i.e. 
\begin{equation}
\Delta_{\pm}(\hat{n},\tau) = \sum_{\ell \, m} a_{\pm2,\,\ell\, m} \, _{\pm 2}Y_{\ell\, m}(\hat{n}).
\label{int2}
\end{equation}
The E- and B-modes are, up to a sign, the real and the imaginary 
parts of $a_{\pm 2,\ell\,m}$, i.e. 
\begin{equation}
a^{(\mathrm{E})}_{\ell\, m} = - \frac{1}{2}(a_{2,\,\ell m} + a_{-2,\,\ell m}), \qquad  
a^{(\mathrm{B})}_{\ell\, m} =  \frac{i}{2} (a_{2,\,\ell m} - a_{-2,\,\ell m}).
\label{int3}
\end{equation}
In real space (as opposed to Fourier space), the fluctuations constructed from 
$a^{(\mathrm{E})}_{\ell\,m}$ and $a^{(\mathrm{B})}_{\ell\,m}$ have the 
property of being invariant under rotations on a plane orthogonal 
to $\hat{n}$.  They can therefore 
be expanded in terms of (ordinary) spherical harmonics:
\begin{equation}
\Delta_{\mathrm{E}}(\hat{n},\tau) = \sum_{\ell\, m} N_{\ell}^{-1} \,  a^{(\mathrm{E})}_{\ell\, m}  \, Y_{\ell\, m}(\hat{n}),\qquad 
\Delta_{\mathrm{B}}(\hat{n},\tau) = \sum_{\ell\, m} N_{\ell}^{-1} \,  a^{(\mathrm{B})}_{\ell\, m}  \, Y_{\ell\, m}(\hat{n}),
\label{int4}
\end{equation}
where $N_{\ell} = \sqrt{(\ell - 2)!/(\ell +2)!}$.  Under parity inversion $a^{(\mathrm{E})}_{\ell\, m} \to (-1)^{\ell}  \,a^{(\mathrm{E})}_{\ell\, m}$ 
while $a^{(\mathrm{B})}_{\ell\, m} \to (-1)^{\ell+1}  \,a^{(\mathrm{B})}_{\ell\, m}$. 
The EE and BB angular power spectra are then defined as 
\begin{equation}
C_{\ell}^{(\mathrm{EE})} = \frac{1}{2\ell + 1} \sum_{m = -\ell}^{\ell} 
\langle a^{(\mathrm{E})*}_{\ell m}\,a^{(\mathrm{E})}_{\ell m}\rangle,\qquad 
C_{\ell}^{(\mathrm{BB})} = \frac{1}{2\ell + 1} \sum_{m=-\ell}^{\ell} 
\langle a^{(\mathrm{B})*}_{\ell m}\,a^{(\mathrm{B})}_{\ell m}\rangle,
\label{int5}
\end{equation}
where $\langle ...\rangle$ denotes the ensemble average. 
In the minimal version of the $\Lambda$CDM paradigm the adiabatic fluctuations of the scalar curvature 
lead to a polarization which is characterized exactly by the condition $a_{2,\,\ell m} = a_{-2,\,\ell m}$, i.e. $a_{\ell m}^{(\mathrm{B})} =0$.  In more general situations 
also the cross-correlations of the various modes can be defined in full terms.
Having introduced the appropriate rotationally invariant observables, the 
empirical observations can now be quickly summarized, in a bulleted form, as\footnote{Following a common 
 practice the polarization will be 
described in terms of the autocorrelations and cross-correlations of the 
E-modes and of the B-modes (i.e., in the jargon, the EE, BB and EB 
angular power spectra). Relevant pieces of information are also encoded 
in the cross-correlation between the temperature and the polarization, i.e. the TE and TB correlations.}:
\begin{itemize}
\item{} the TE and the EE angular power spectra have been measured so far by various experiments with different accuracies;
\item{} direct measurements of the BB angular power spectra led just 
upper limits on the B-mode polarization since the observations are consistent with the null  result;
\item{} the polarization power spectra detected by different experiments are consistent.
\end{itemize}
\begin{table}[!ht]
\begin{center}
\begin{tabular}{|c|c|c|c|c|}
\hline
Experiments & Years  & Data & Pivot frequencies &Upper limits on BB \\
\hline
Dasi \cite{dasi1,dasi2,dasi3} & 2000-2003 & EE, TE, BB &
$26$--$36$ GHz & ${\mathcal G}_{\ell}^{(\mathrm{BB})} \leq 50 \,(\mu\,\mathrm{K})^2$  
\\
Boomerang \cite{boom1}&2003  & EE, EB, BB & $145$ GHz& ${\mathcal G}_{\ell}^{(\mathrm{BB})}\leq 8.6 \,(\mu\,\mathrm{K})^2$ 
 \\
Maxipol \cite{maxip1,maxip2} & 2003 & EE, EB, BB & $140$ GHz & 
${\mathcal G}_{\ell}^{(\mathrm{BB})} \leq 112.3 \,(\mu\,\mathrm{K})^2$ 
\\
Quad \cite{quad1,quad2,quad3} & 2007 & EE,TE, TB& $100$/ $150$ GHz&
${\mathcal G}_{\ell}^{(\mathrm{BB})} \leq  10 \,(\mu\,\mathrm{K})^2$
\\
Cbi \cite{cbi} & 2002-2005& EE, TE, BB  & $26$--$36$ GHz& 
${\mathcal G}_{\ell}^{(\mathrm{BB})} \leq  3.76\,(\mu\,\mathrm{K})^2$
\\
Capmap  \cite{capmap} & 2004-2005& EE, BB& $35$--$46$/$84$--$100$ GHz& 
${\mathcal G}_{\ell}^{(\mathrm{BB})} \leq  4.8\,(\mu\,\mathrm{K})^2$
\\
Wmap \cite{WMAPfirst1,WMAP51} & 2001-2006& EE, TE, BB& $23$--$94$ GHz& 
${\mathcal G}_{\ell}^{(\mathrm{BB})} \leq  0.15\,(\mu\,\mathrm{K})^2$
\\
\hline
\end{tabular}
\caption{The main polarization experiments, their typical frequencies, their
salient observations are illustrated 
in the light of the polarization properties of the CMB. In the last column 
${\mathcal G}_{\ell}^{(\mathrm{BB})}=\ell (\ell +1)C_{\ell}^{(\mathrm{BB})}/(2\pi)$.}
\label{TABLE1}
\end{center}
\end{table}
In Tab. \ref{TABLE1} the salient informations on different experiments measuring the polarization properties of the CMB are swiftly collected. 
In Tab. \ref{TABLE1}, the last two columns at the right include specific
 informations 
on the frequency channels characteristic of each experiment and of the upper limits on the B-mode autocorrelation.  
The frequency of the observational channel constitutes a relevant piece 
of information since, 
as already remarked, the Faraday rate depends upon the 
wavelength of the propagating radiation. 
The main references to each experiment have been  collected in the first column of  Tab. \ref{TABLE1}. In the case of the WMAP experiment 
the most recent data release can be found in \cite{WMAP51,WMAP52,WMAP53,WMAP54,WMAP55}. The first data release corresponds to 
\cite{WMAPfirst1,WMAPfirst2,WMAPfirst3}. For the specific aspects of the polarization measurements it is also interesting to bear in mind Ref. \cite{WMAPthird2} (see also \cite{WMAPthird1}).

The paradigm in vogue these days is the $\Lambda$CDM scenario 
where $\Lambda$ stands for the dark energy component 
and CDM for the cold dark matter component.
 In the $\Lambda$CDM context the origin of the linear polarization 
of the CMB stems directly from the properties of last scattering in conjunction with the presence of a primordial spectrum of curvature perturbations which is said to be adiabatic. Adiabatic means, in the present context, that 
the curvature perturbations do not arise as inhomogeneities in the plasma sound speed 
but, on the contrary, they are related to genuine fluctuations of the energy density and of the spatial curvature.
In the adiabatic case the location of the  first acoustic peak of the TT angular power spectrum determines 
 the location of the first anticorrelation peak of the 
TE spectrum. This is what is observed since, roughly, $\ell_{\mathrm{TT}} \simeq 220$ 
and $\ell_{\mathrm{TE}} \simeq 3 \ell_{\mathrm{TT}}/4$ where $\ell_{\mathrm{TT}}$ denotes the location of the first acoustic peak 
and $\ell_{\mathrm{TE}}$ corresponds to the first anticorrelation peak 
in the TE spectrum.  Finally the success of the 
$\Lambda$CDM scenario in pinning down the observed features in the TT and EE  angular power spectra gives us also, indirectly, some precious informations on the dynamics of photon decoupling and on the 
 possibility of late-time reionization.  As indicated by Tab. \ref{TABLE1} various experiments 
measured (and are now measuring) the EE and TE angular power spectra. On top of the 5-yr WMAP results
\cite{WMAP51,WMAP52,WMAP53,WMAP54,WMAP55} recently there have been rather promising 
results coming from the Quad collaboration \cite{quad2} (see also \cite{quad1,quad3}) claiming 
rather intriguing measurements of the EE and TE angular power spectra up to rather large 
harmonics\footnote{The given measured multipole $\ell$ is related to the angular 
separation. Large multipoles correspond to small angular separations, i.e., roughly $\vartheta = \pi/\ell$.}, 
up to $\ell \simeq 1500$ and even $\ell \simeq 2000$. 

In the $\Lambda$CDM 
scenario with no tensors, the B-modes vanish  except for a minute component coming from the gravitational lensing of the primary anisotropies. If the $\Lambda$CDM scenario is complemented 
by means of a tensor contribution, then the value of the BB autocorrelations 
will depend upon the ratio between the power spectrum of the tensor 
fluctuations and the power spectrum of the fluctuations of the scalar 
curvature.
\begin{figure}[!ht]
\centering
\includegraphics[height=8cm]{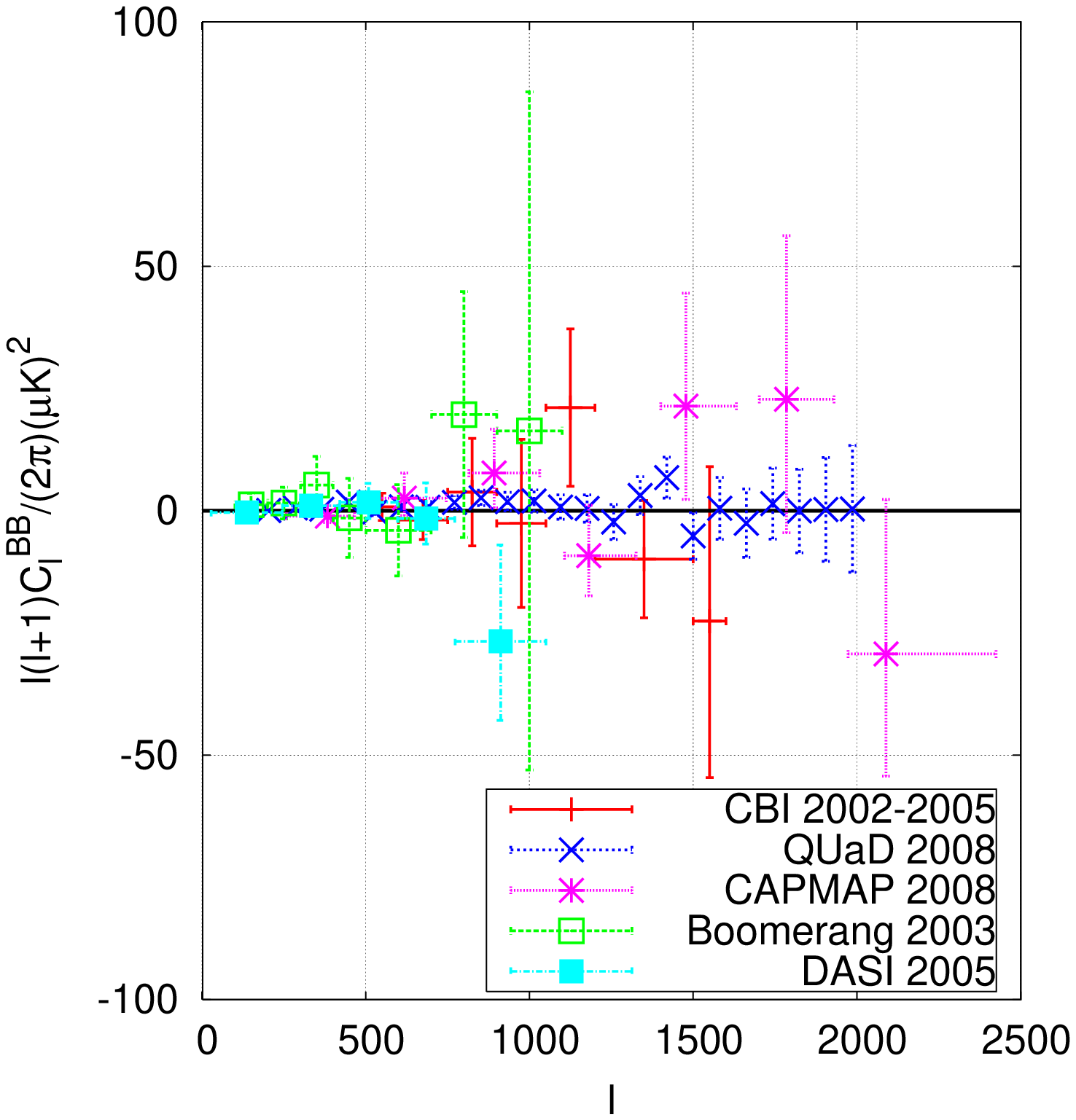}
\includegraphics[height=8cm]{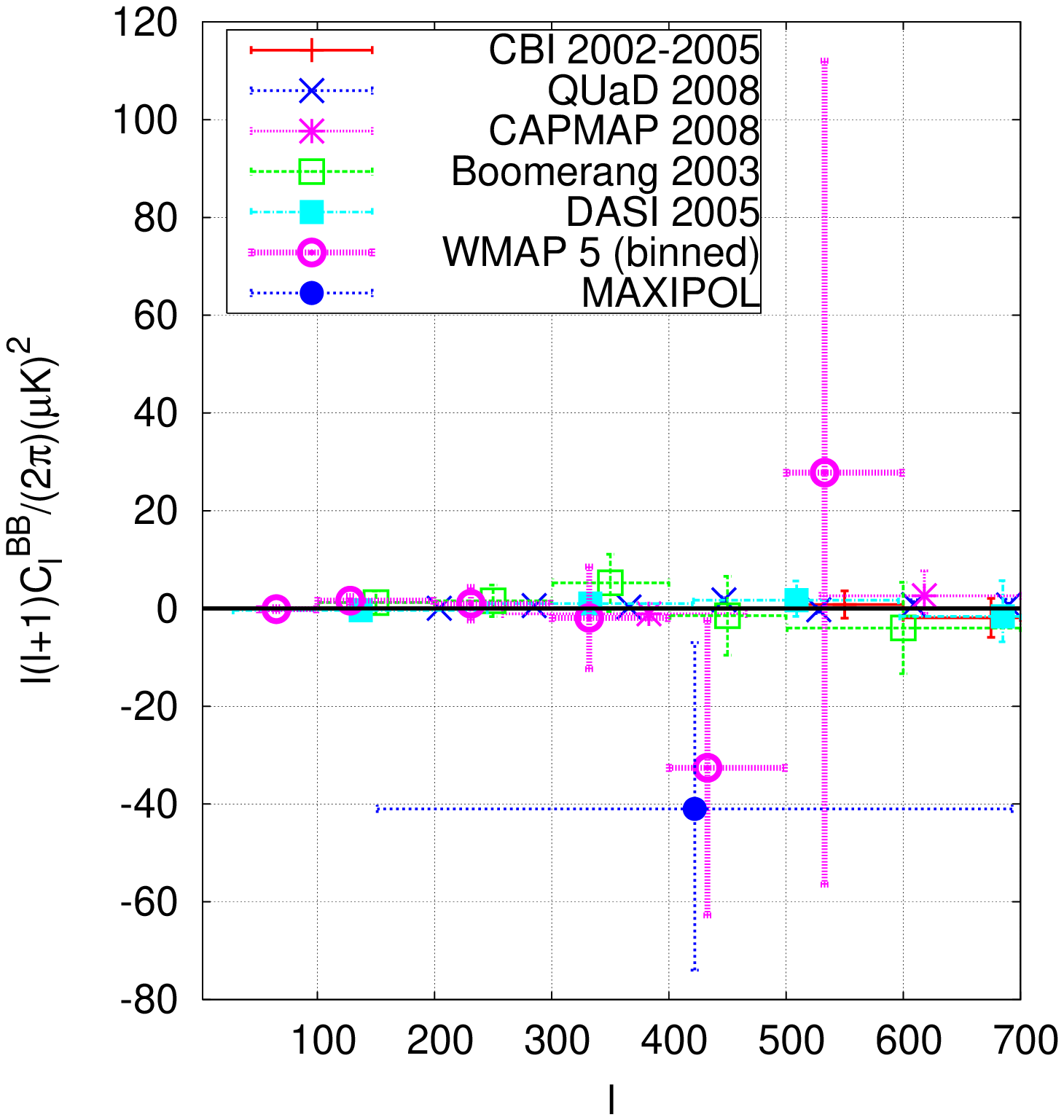}
\caption[a]{Measurements of the BB autocorrelations as they have been reported in 
the various experiments listed in Tab. \ref{TABLE1}.}
\label{Figure1}      
\end{figure}
The current experimental data reported in 
Tab. \ref{TABLE1} are also illustrated in 
Fig. \ref{Figure1}. While the possible presence of a B-mode polarization can be attributed to a stochastic 
background of relic gravitons, Eqs. (\ref{heat1})--(\ref{heat3}) 
 suggest that it could be the signal of a pre-decoupling magnetic field.
 
Equation (\ref{heat1}) describes the evolution of the intensity of the radiation field and, in a magnetized medium, the source term ${\mathcal S}_{I}$ receives contribution from both the fluctuations of the geometry as well as from the fluctuations of the plasma which is, in turn, sensitive to the stochastic magnetic field itself.
The remaining two equations determine the polarization observables and can be solved by iteration, i.e. by positing that 
\begin{equation}
Q = \overline{Q} + Q^{(1)},\qquad U= \overline{U} + U^{(1)}.
\label{heat5}
\end{equation}
The first iteration determines $\overline{Q}$ which is a solution of Eqs. (\ref{heat1})--(\ref{heat2}) with $\overline{U} =0$, i.e.  
\begin{equation}
\frac{d \overline{Q}}{d\tau} + \epsilon' \overline{Q} = \epsilon' {\mathcal S}_{\overline{Q}}, \qquad 
\frac{d U^{(1)}}{d\tau} + \epsilon' U^{(1)} =  - 2 \epsilon' \kappa_{\mathrm{F}} \overline{Q}.
\label{heat6}
\end{equation}
The result of the second iteration produces $U^{(1)}\neq 0$ and, ultimately, a B-mode polarization.
In the present paper the magnetized B-mode will be computed both analytically and numerically. 
Furthermore various bounds on the magnetic field intensity will be derived 
by comparing the  Faraday-induced B-mode both
with the upper limits on the BB angular power spectra (see  Tab. \ref{TABLE1}) 
as well as with the published data on the B-mode angular power spectrum (see e. g. Fig. \ref{Figure1}). 

The present investigation is based on a number of previous results. 
The large-scale magnetic fields are included both at the level 
of the evolution equations as well as at the level of the initial conditions. This program has been undertaken  in \cite{mg1} (see also \cite{prev,maxrev1}) and subsequently developed in \cite{mg2,mg3}.
Semi-analytic results for the polarization observables which are relevant for 
the present investigation have been discussed in \cite{mg2}.
Recently the results of \cite{mg1,mg2,mg3}  have been complemented by 
a dedicated numerical approach \cite{gk1,gk2,gk3} which 
is based, indirectly, on COSMICS \cite{cosmics1,cosmics2} and, more directly, on CMBFAST \cite{cmbfast1,cmbfast2}.
 
In \cite{gk3} a bound on pre-decoupling magnetic fields has been derived, by including, as suggested in \cite{mg2}, the large-scale magnetic fields 
in the heat transfer equations as well as in the initial conditions of the full Einstein-Boltzmann 
hierarchy.   The obtained results 
showed that the bounds obtainable from the TT correlations were still more constraining 
than the direct Faraday bounds stemming from the upper limits on the B-mode autocorrelations. 
The assumptions of \cite{gk3} will be relaxed and the 
finite thickness of the last scattering surface will be taken into account in the analytical and numerical 
estimate of the Faraday autocorrelation.
While the sudden decoupling limit (i.e. infinitely thin last-scattering surface) 
is  justified for large angular scales (i.e. $\ell < 100$) over small angular scales finite thickness effects 
are comparable with diffusive effects.  The latter statements can be understood 
also in analogy with the calculations of the temperature autocorrelations.
While the finite thickness 
effects can be ignored for the estimate of the Sachs-Wolfe plateau, the same 
cannot be said for the region of the acoustic peaks where both the finite thickness 
effects compete with diffusive damping to determine the well known 
shape of the temperature autocorrelations. Semi-analytic  treatments of the 
finite thickness effects can be found in  \cite{zal,wyse,zeld1,pav1,pav2}. 
Finally, finite thickness effects also determine the correlated distortions 
of the first three acoustic peaks  in the temperature autocorrelations when 
 large-scale magnetic fields are consistently included in the calculation \cite{mg3,gk2}.
 
 In the light of some of the recent (and forthcoming) experimental data,
 finite thickness effects play a relevant role in the estimate of the Faraday autocorrelations as well as in the estimate of the B-mode autocorrelations. Indeed (see Tab. \ref{TABLE1} and Fig. \ref{Figure1}) the 
upper limits on the B-mode autocorrelations stem from rather small angular scales. Over those scales the Faraday induced B-mode cannot be trusted in the sudden decoupling limit.  
 
While physically the small angular scales are the realm of diffusive effects, geometrically 
the small angular separations imply that the microwave sky  degenerates 
into a plane. On a 2-sphere the angular eigenfunctions of the Laplacian are  spherical harmonics. When the radius of the sphere is taken to infinity (i.e. the angular separation becomes small) the eigenfunctions of the 
Laplacian become plane waves. In the language of group representations this is one 
of the simplest examples of group contraction (in the In\"on\"u-Wigner sense) where the group $SO(3)$ contracts to $M(2)$, i.e. the semidirect product of the Euclidean group in two dimension and of the 
$SO(2)$ group \cite{vilenkin}.  This observation can simplify the 
evaluation of complicated correlators and will lead, ultimately, to an
analytical expression of the Faraday-induced autocorrelations at small angular scales.

The plan of the present investigation is the following. 
In section 2 the essentials of the interplay between large-scale magnetism and CMB 
physics will be reviewed. 
Faraday  autocorrelations will be computed in section 3.  
The B-mode autocorrelations will computed in section 4. 
In Section 5 the parameters of the magnetized $\Lambda$CDM model will 
be analyzed within a frequentist approach. 
The concluding remarks are collected in section 6.
\renewcommand{\theequation}{2.\arabic{equation}}
\setcounter{equation}{0}
\section{Pre-decoupling plasma and large-scale magnetic fields}
\label{sec2}
Heeding observations,  the linear polarization of the CMB is a well established empirical notion. 
Conversely the optical activity of the electron-baryon plasma is still under active scrutiny \cite{maxrev1}. The program suggested in \cite{maxrev1} 
has been to include consistently large-scale 
magnetic fields in all the steps leading to the angular power spectra of the CMB anisotropies. One of the hopes of the program will be 
to gauge accurately the effects of large-scale 
magnetic fields and to set limits on their putative existence. The essentials of the theoretical framework  will now be 
schematically introduced. More detailed derivations 
can be found in \cite{mg1,mg2,mg3,gk1,gk3}. 
The present approach to pre-decoupling magnetism is rooted in three plausible assumptions:
\begin{itemize}
\item{[i]}  large-scale magnetic fields, if present, do not break 
the spatial isotropy and, therefore, cannot be uniform;
\item{[ii]} the curved-space effects as well as the relativistic fluctuations 
of the geometry are described within the lore provided by General Relativity;
\item{[iii]} the same physical laws describing the dynamics of electromagnetic fields in terrestrial plasmas are fully valid.
\end{itemize} 
The implications of each of the three aforementioned hypotheses 
will now be briefly introduced with particular attention to the tools which 
will be exploited in the subsequent part of the present analysis. 
The necessity of leaving unbroken spatial isotropy demands that the Fourier modes of the magnetic field must obey 
\begin{equation}
\langle B_{i}(\vec{k}, \tau) B_{j}(\vec{p},\tau) \rangle = \frac{2\pi^2}{k^3}\,\, {\mathcal P}_{\mathrm{B}}(k) P_{ij}(k) 
 \delta^{(3)}(\vec{k} + \vec{p}), 
\label{EQ1}
\end{equation}
where
\begin{equation}
P_{ij}(k) = \biggl(\delta_{ij} - \frac{k_{i} k_{j}}{k^2}\biggr),\qquad P_{\mathrm{B}}(k) = A_{\mathrm{B}} 
\biggl(\frac{k}{k_{\mathrm{L}}}\biggr)^{n_{\mathrm{B}}-1}.
\label{EQ2}
\end{equation}
In Eq. (\ref{EQ2}) $A_{\mathrm{B}}$ is the spectral amplitude evaluated 
at the magnetic pivot scale $k_{\mathrm{L}}$; $n_{\mathrm{B}}$ is the magnetic spectral index. Within the conventions of Eqs. (\ref{EQ1}) and (\ref{EQ2}), $A_{\mathrm{B}}$ will have dimensions of an energy density and will therefore be measured in $\mathrm{nG}^2$. 
The scale-invariant spectrum of magnetic inhomogeneities does correspond to 
$n_{\mathrm{B}} =1$.  In Eqs. (\ref{EQ1})--(\ref{EQ2}) 
the correlation functions have been assigned within the same conventions employed to 
assign the curvature perturbations on comoving orthogonal hypersurfaces, i.e. 
\begin{equation}
\langle {\mathcal R}_{*}(\vec{k}) {\mathcal R}_{*}(\vec{p}) \rangle = \frac{2\pi^2}{k^3} {\mathcal P}_{{\mathcal R}}(k) \delta^{(3)}(\vec{k}+ \vec{p}), \qquad {\mathcal P}_{{\mathcal R}} = {\mathcal A}_{{\mathcal R}} \biggl(\frac{k}{k_{\mathrm{p}}}\biggr)^{n_{\mathrm{s}} -1}.
\label{EQ2a}
\end{equation}
In Eq. (\ref{EQ2a}) ${\mathcal A}_{\mathcal R}$ denotes the amplitude of curvature perturbations 
at the pivot scale $k_{\mathrm{p}}= 0.002\,\, \mathrm{Mpc}^{-1}$.  
Owing to the isotropy (and spatial flatness) of the background geometry,
the line element  can be written, in the conformally flat parametrization, as
\begin{equation}
ds^2 = a^2(\tau) [ d\tau^2 - d\vec{x}^2].
\label{EQ3}
\end{equation}
The scale factor will always be normalized in such a way that $a_{0}=1$.
With the latter convention 
comoving and physical scales coincide at the present time.
This convention will be enforced both in the analytic calculations 
as well as in the numerical applications. 

Using Eq. (\ref{EQ3}), the second hypothesis listed above (i.e. [ii])
implies that $a(\tau)$ obeys 
the Friedmann-Lema\^itre equations which can be written as 
\begin{eqnarray}
&&{\mathcal H}^2 = \frac{8\pi G}{3} a^2 \rho_{\mathrm{t}}, \qquad {\mathcal H}^2 - {\mathcal H}' = 4 \pi G a^2 (p_{\mathrm{t}} + \rho_{\mathrm{t}}), 
\label{FL1}\\
&& \rho_{\mathrm{t}}' + 3 {\mathcal H} (\rho_{\mathrm{t}} + p_{\mathrm{t}}) =0,\qquad {\mathcal H} = \frac{a'}{a},
\label{FL2}
\end{eqnarray}
where $p_{\mathrm{t}}$ and $\rho_{\mathrm{t}}$ denote, respectively, the total pressure and the total energy density, i.e. 
\begin{equation}
\rho_{\mathrm{t}} = \rho_{\mathrm{e}} + \rho_{\mathrm{i}} + \rho_{\nu} + \rho_{\gamma}
+ \rho_{\mathrm{c}} + \rho_{\Lambda}.
\label{FL3}
\end{equation}
The various components of Eq. (\ref{FL3}) refer, respectively, to the electrons, the ions, the 
neutrinos, the photons, the cold dark matter (CDM) species, and the dark energy.  The relativistic fluctuations of the geometry will be  described 
within the longitudinal gauge where the perturbed entries 
of the metric take the form:
\begin{equation}
\delta_{\mathrm{s}} g_{00} = 2 a^2 \phi(\vec{x},\tau),\qquad \delta_{\mathrm{s}} g_{ij} = 2 a^2 \psi(\vec{x},\tau) \delta_{ij}.
\label{EQ4}
\end{equation}
As previously stressed (see, e.g. \cite{mg1,mg2,gk2}) 
the longitudinal gauge is particularly suitable to carry on semi-analytic 
estimates. Conversely the synchronous coordinate 
system turns out to be rather handy in dedicated numerical calculations. 
The subtleties related to the use of different coordinate 
systems in the description of magnetized plasmas have been 
thoroughly explored in the recent past \cite{prev} (see also \cite{gk2}). 

From the third (and last) assumption listed at the beginning of this 
section the evolution of the electromagnetic fields follows, schematically,  Maxwell's equations
\begin{eqnarray}
&&\vec{\nabla}\cdot \vec{E} = 4\pi e \int d^{3}v [f_{+}(\vec{x}, \vec{v}, \tau) - f_{-}(\vec{x}, \vec{v}, \tau)], \qquad \vec{\nabla}\cdot \vec{B} =0,
\label{MX1}\\
&& \vec{\nabla} \times \vec{E}  + \vec{B}' = 0, \qquad \vec{\nabla}\times \vec{B} - \vec{E}'= 
4\pi e \int d^{3}v\,\vec{v}\,[f_{+}(\vec{x}, \vec{v}, \tau) - f_{-}(\vec{x}, \vec{v}, \tau)],
\label{MX2}
\end{eqnarray}
where $f_{\pm}(\vec{x}, \vec{v}, \tau)$ are the one-body distribution functions for electrons and ions written in terms of the comoving velocity $\vec{v}$ (and not upon the comoving three-momentum $\vec{q}$) since, after neutrino decoupling, electrons and ions are non-relativistic, i.e. $m_{\mathrm{e}} \gg T_{\mathrm{e}}$ 
and $m_{\mathrm{i}} \gg T_{\mathrm{i}}$ where $T_{\mathrm{e}}$ and 
$T_{\mathrm{i}}$ denote, respectively, the electron and ion 
temperatures which are, in the first approximation, of the order 
of the temperature of the photons.  Consequently the 
comoving three-velocity and the comoving three-momentum 
are related as  
\begin{equation}
\vec{v}_{\mathrm{e},\mathrm{i}} = \frac{\vec{q}_{\mathrm{e},\mathrm{i}}}{\sqrt{q_{\mathrm{e}\,\mathrm{i}}^2 + m_{\mathrm{e}\,\mathrm{i}}^2 a^2}} \simeq  
\frac{\vec{q}_{\mathrm{e},\mathrm{i}}}{m_{\mathrm{e}\,\mathrm{i}} a},
\label{VEL1}
\end{equation}
where the subscripts refer, indifferently either to electrons or to ions. 
In Eq. (\ref{VEL1}) the relativistic fluctuations of the geometry 
have been neglected just to avoid complications. In explicit 
calculations, however, they must be taken into account. Indeed, 
the one-body distribution functions will obey the corresponding Vlasov-Landau equations, i.e.
\begin{equation}
\frac{\partial f_{\pm}}{\partial \tau} + v^{i} \frac{\partial f_{\pm}}{\partial x^{i}} \pm e 
( E^{i} + v_{j} B_{k} \epsilon^{j\,k\,i}) \frac{\partial f_{\pm}}{\partial q^{i}}+  q^{j}[\psi' - n_{a} \partial^{a} \phi] \frac{\partial f_{\pm}}{\partial q^{j}} = 
{\mathcal C}_{\mathrm{coll}},
\label{VL}
\end{equation}
where $\psi$ and $\phi$ are the two entries of the perturbed metric introduced in Eq. (\ref{EQ4}) and ${\mathcal C}_{\mathrm{coll}}$ is the collision term which is dictated by electron-ion scattering as well
as by the scattering with the photons whose concentration, around decoupling, is $10^{10}$ 
times larger than the one of electrons and ions \cite{gk3}. 
The electromagnetic fields appearing in Eqs. (\ref{MX1})--(\ref{MX2}) are  all comoving but, as it can be appreciated from Eq. (\ref{VEL1}) 
the presence of electrons and ions breaks explicitly 
the conformal invariance of the whole system. In particular 
this will reflect, as discussed in the past, in the nature 
of the tight-coupling approximation \cite{prev}.

Equations (\ref{MX1}), (\ref{MX2}) and (\ref{VL}) summarize, in a nutshell,  
some of the salient features of the description 
of electromagnetic plasmas in curved space-times and in the 
presence of the relativistic fluctuations of the geometry. 
At early times (i.e. well prior to matter-radiation equality) 
the Boltzmann-hierarchy must be treated by making 
explicit the tight Coulomb coupling as well as the 
tight Thomson coupling.  Later on this approximation will break down.
The general approximation scheme will now be outlined leaving aside  
details which can be however found elsewhere.

Given the minuteness of the Debye 
scale\footnote{The Debye scale has, in the 
present context, exactly the same meaning and the same form 
it would have, for instance, in a terrestrial plasma, i.e. 
$\lambda_{\mathrm{D}} = 
\sqrt{T_{\mathrm{e}}/(8\pi \, e^2 \tilde{n}_{\mathrm{e}})}$. For typical lenght-scales $L > \lambda_{\mathrm{D}}$ the collective effects 
dominate and the positive charge carriers are compensated by the 
negative charge carriers. In the opposite limit the propagation 
of electromagnetic disturbances  cannot be neglected.} in comparison with the Hubble radius the plasma approximation works extremely well. A two fluid description can then  be employed 
to describe the electron-ion system before decoupling. Taking the first moment of Eq. (\ref{VL}) we can get the evolution of the comoving velocity fields for electrons (i.e. $\vec{v}_{\mathrm{e}}$) and for ions (i.e. $\vec{v}_{\mathrm{i}}$):
\begin{eqnarray}
&&\vec{v}_{\mathrm{e}}\,' + {\mathcal H} \vec{v}_{\mathrm{e}} = - 
\frac{e n_{\mathrm{e}}}{\rho_{\mathrm{e}} a^4} [\vec{E} + \vec{v}_{\mathrm{e}}\times \vec{B}] - \vec{\nabla}\phi  + {\mathcal C}_{\mathrm{e\,i}} + {\mathcal C}_{\mathrm{e}\,\gamma},
\label{EQ5}\\
&&\vec{v}_{\mathrm{i}}\,' + {\mathcal H} \vec{v}_{\mathrm{i}} =  
\frac{e n_{\mathrm{i}}}{\rho_{\mathrm{i}} a^4} [\vec{E} + \vec{v}_{\mathrm{i}}\times \vec{B}]  - \vec{\nabla} \phi + {\mathcal C}_{\mathrm{i\,e}} + {\mathcal C}_{\mathrm{i}\,\gamma},
\label{EQ6}
\end{eqnarray}
where $n_{\mathrm{e}} = a^3\tilde{n}_{e}$ and $n_{\mathrm{i}} = a^3\tilde{n}_{\mathrm{i}}$.  
In Eq. (\ref{EQ5})  ${\mathcal C}_{\mathrm{e\,i}}$ 
and ${\mathcal C}_{\mathrm{e}\,\gamma}$ denote, respectively,  the collision 
terms between electrons and ions and the collision terms between 
electrons and photons. With analog notation the collision terms 
of the ions are included in Eq. (\ref{EQ6}). The electron-photon 
cross section dominates the mean free path of the photons
because of the hierarchy between the electron and proton masses.

For comoving scales much larger than the (comoving) Debye scale $\lambda_{\mathrm{D}}$ and for frequencies much smaller than the plasma frequency  the electromagnetic fields obey an effective one-fluid 
description which arises naturally by assuming that the Coulomb coupling between electrons 
and ions is strong \cite{gk3}. In this limit the effective set of equations is  given by the 
generalization of magnetohydrodynamics (MHD) to the case when the geometry is curved and it fluctuates 
according to Eq. (\ref{EQ4}). This set of equations has been written in 
\cite{mg2} and will not be repeated here. The evolution equation 
of the electron-ion velocity can be directly obtained by carefully summing up Eqs. (\ref{EQ5}) and 
(\ref{EQ6}).  After simple algebra, the result is\footnote{Typically $m_{\mathrm{i}} \simeq m_{\mathrm{p}}$ where $m_{\mathrm{p}}$ is the 
proton mass.}:
\begin{equation}
\theta_{\mathrm{b}}' + {\mathcal H} \theta_{\mathrm{b}} = \frac{\vec{\nabla} \cdot [ \vec{J} \times \vec{B}]}{\rho_{\mathrm{b}}a^4 \biggl( 1 + \frac{m_{\mathrm{e}}}{m_{\mathrm{p}}}\biggr)} - \nabla^2\phi + \frac{4}{3} \frac{\rho_{\gamma}}{\rho_{\mathrm{b}}} \epsilon' (\theta_{\gamma} - \theta_{\mathrm{b}}),
\label{EQ7}
\end{equation}
where $\theta_{\gamma} = \vec{\nabla}\cdot\vec{v}_{\gamma}$  is the three-divergence 
of the comoving three-velocity of the photons and where\footnote{The absence of the ions from the photon mean free path (and in the corresponding 
rate) is due to the fact that the ions contribute little to the photon mean free path since their cross section 
is negligible in comparison with the one of the electrons.}:
\begin{eqnarray}
\epsilon' &=& \tilde{n}_{\mathrm{e}} x_{\mathrm{e}} \sigma_{\mathrm{Th}} a,\qquad \sigma_{\mathrm{Th}} = \frac{8}{3} \pi \biggl(\frac{e^2}{m_{\mathrm{e}}}\biggr)^2,
\label{EQ8}\\
\theta_{\mathrm{b}} &=& \frac{m_{\mathrm{e}} \theta_{\mathrm{e}} + m_{\mathrm{p}} \theta_{\mathrm{i}}}{m_{\mathrm{e}} + m_{\mathrm{i}}}, \qquad\theta_{\mathrm{e}}= \vec{\nabla}\cdot \vec{v}_{\mathrm{e}}, 
\qquad \theta_{\mathrm{i}}= \vec{\nabla}\cdot \vec{v}_{\mathrm{i}}.
\label{EQ9}
\end{eqnarray}
In Eq. (\ref{EQ7}), $\vec{J}$ denotes the (solenoidal) Ohmic current. Notice that $\theta_{\mathrm{b}}$ is effectively the centre-of mass velocity of the electron-ion system; $\theta_{\mathrm{b}}$ in the  limit 
of strong Coulomb coupling $\theta_{\mathrm{b}}$ coincides 
with the bulk velocity of the plasma.
Equation (\ref{EQ7}) is coupled directly both to the evolution 
of the density contrasts as well as to the evolution of the photon 
velocity field, i.e. 
\begin{eqnarray}
&& \delta_{\gamma}' + \frac{4}{3} \theta_{\gamma} - 4 \psi' =0,
\label{DC1}\\
&& \delta_{\mathrm{b}}' + \theta_{\mathrm{b}} - 3 \psi' = \frac{\vec{E}\cdot\vec{J}}{a^4 \rho_{\mathrm{b}} \biggl(1 + \frac{m_{\mathrm{e}}}{m_{\mathrm{p}}}\biggr)},
\label{DC2}\\
&& \theta_{\gamma}' + \frac{1}{4} \nabla^2 \delta_{\gamma} + \nabla^2 \phi = \epsilon' (\theta_{\mathrm{b}} - \theta_{\gamma}).
\label{DC3}
\end{eqnarray}
Since 
the electron-ion system is strongly coupled also to photons it is in principle possible to consider 
photons, baryons and electrons as a unique dynamical entity \cite{mg2}:
\begin{equation}
\theta_{\gamma\mathrm{b}}' + {\mathcal H} \frac{R_{\mathrm{b}}}{R_{\mathrm{b}} + 1} \theta_{\gamma\mathrm{b}} = \frac{3}{4 \rho_{\gamma} a^4 \,( 1 + R_{\mathrm{b}})} \vec{\nabla}\cdot [\vec{J} \times \vec{B}] - 
\frac{\nabla^2 \delta_{\gamma}}{4 ( 1 + R_{\mathrm{b}})} - \frac{\nabla^2\phi}{1 + R_{\mathrm{b}}},
\label{EQ10}
\end{equation}
where\footnote{Following the usual notations 
$\Omega_{\mathrm{b}0} = \rho_{\mathrm{b}0}/\rho_{\mathrm{crit}}$ 
denotes the critical fraction of baryons at the present time. For a generic 
component $x$ we will have that 
$\Omega_{x\,0} = \rho_{x\,0}/\rho_{\mathrm{crit}}$.}  $R_{\mathrm{b}}(\tau)$ 
\begin{equation}
R_{\mathrm{b}}(\tau) = \frac{3}{4} \frac{\rho_{\mathrm{b}}}{\rho_{\gamma}} = 690.18 \biggl(\frac{h_{0}^2 \Omega_{\mathrm{b}0}}{0.02273}\biggr) a(\tau),
\label{EQ11}
\end{equation}
is the baryon-to-photon ratio. Equation (\ref{EQ10}) can be easily derived 
by combining Eqs. (\ref{EQ7}) and (\ref{DC3}) and by noticing that when 
the photon-electron-baryon coupling is tight $\theta_{\mathrm{b}}
\simeq \theta_{\gamma} \simeq \theta_{\gamma\mathrm{b}}$. 
By going from Eq. (\ref{VL}) 
to Eq. (\ref{EQ11}) the effective number of degrees of freedom has been progressively reduced. The effective description mentioned so far is very useful at early times and over 
large scales (i.e. small multipoles). It should not be applied, however, outside its domains of validity.

The breakdown of the effective one-fluid 
description is related to the possible propagation of electromagnetic 
disturbances. Indeed, in MHD the Ohmic current is solenoidal (i.e. $4\pi \vec{J} = 
\vec{\nabla}\times \vec{B}$) and 
the electric fields vanish in the baryon rest frame because 
of the large value of the conductivity \cite{mg1}. The latter statement depends upon the frequency of the electromagnetic radiation. It is actually  
well known that high-frequency waves can propagate inside (magnetized)
globally neutral plasmas \cite{alfven,krall}. 

The baryon-photon system ceases to be a unique 
physical entity around the decoupling time. As time goes by 
the velocity fields of the photons and of the baryons must be separately followed by numerical means. The electron-ion system 
does not break down after a particular time but above a particular frequency of the spectrum of plasma excitations. 
If the plasma is not magnetized the only typical scale 
of the problem is set by the electron and ion concentrations which determine 
the corresponding (comoving) plasma frequencies, i.e. 
\begin{eqnarray}
&& \overline{\omega}_{\mathrm{pe}} = \sqrt{\frac{4\pi n_{0} x_{\mathrm{e}} e^2}{m_{\mathrm{e}} a}}  
\equiv  \omega_{\mathrm{pe}} a,
\qquad \omega_{\mathrm{pe}} =\sqrt{\frac{4\pi \tilde{n}_{\mathrm{e}} x_{\mathrm{e}} e^2}{m_{\mathrm{e}}}},
\label{EQ12}\\
&& \overline{\omega}_{\mathrm{pi}} = \sqrt{\frac{4\pi n_{0} x_{\mathrm{e}} e^2}{m_{\mathrm{p}} a}}  
\equiv  \omega_{\mathrm{pi}} a,
\qquad \omega_{\mathrm{pi}} =\sqrt{\frac{4\pi \tilde{n}_{\mathrm{i}} x_{\mathrm{i}} e^2}{m_{\mathrm{p}}}},
\label{EQ13}
\end{eqnarray}
where it has been used that $n_{0} = a^{3}\, \tilde{n}_{\mathrm{e}}= a^3 \tilde{n}_{\mathrm{i}}$.
Taking into account that the comoving concentration of charge carriers is
$10^{10}$ times smaller than the one of photons
\begin{equation}
n_{0} = \eta_{\mathrm{b}0} n_{\gamma0},\qquad \eta_{\mathrm{b}0} = 6.219 \times 10^{-10} \biggl(\frac{h_{0}^2\Omega_{\mathrm{b}0}}{0.02273}\biggr) \biggl(\frac{T_{\gamma 0}}{2.725}\biggr)^{-3}, 
\label{EQ14}
\end{equation}
we can say that the effective bulk description provided by Eq. (\ref{EQ7}) and its descendants break down for physical frequencies larger than \cite{mg3}
\begin{equation}
\omega_{\mathrm{pe}}(z) = 0.0285 \, \sqrt{x_{\mathrm{e}}} \, \biggl( \frac{h_{0}^2 \Omega_{\mathrm{b}0}}{0.02273}\biggr)^{1/2} \biggl(\frac{z +1}{1000}\biggr)\,\, \mathrm{MHz}, \qquad \omega_{\mathrm{pi}}(z) = \sqrt{\frac{m_{\mathrm{e}}}{m_{\mathrm{p}}}} \,\omega_{\mathrm{pe}}(z).
\label{EQ15}
\end{equation}
As already stressed in \cite{gk3}  to treat the dispersive propagation of electromagnetic waves 
in a plasma demands necessarily that electrons are treated separately from the ions. The second point to be  emphasized is that the inhomogeneity scale of the magnetic fields is typically much larger than the Larmor radius. This hierarchy of scales justifies 
the use of the guiding centre approximation \cite{alfven,krall} in the derivation of the relevant dispersion relations.  The guiding centre  
approximation is well justified since, in the problem at hand, the 
scale of variation of the magnetic field is (always) much larger 
than the gyration radius for ions and electrons. 

The spectrum of plasma excitations may be in principle very rich
and it includes a load of different phenomena which have both 
practical and conceptual implications \cite{stix}. 
The dispersive propagation associated with the ordinary and extraordinary wave is, however, irrelevant in the case of the CMB \cite{gk3}. 
In similar terms, other dispersive effects are also absent.
The rationale for these occurrences stems from the hierarchy between 
the plasma (and Larmor) frequencies of electrons and ions (see e.g. Eq. (\ref{EQ15})) and the frequency of the maximum of the CMB which is 
given by
\begin{equation}
\nu_{\mathrm{max}} = \frac{x_{\mathrm{max}}}{2\pi} \,T_{\gamma 0}= 222.617 \biggl(\frac{T_{\gamma 0}}{2.725\, \mathrm{K}}\biggr)\, \mathrm{GHz},\qquad x_{\mathrm{max}} = 3.920,
\label{rot11a}
\end{equation}
where $x= k/T_{\gamma 0}$ and $x_{\mathrm{max}}$ corresponds to the maximum of the spectral energy density of the CMB, i.e. 
\begin{equation}
\Omega_{\gamma}(x) = \frac{15}{\pi^4} \Omega_{\gamma0} \frac{x^4}{e^{x} -1},\qquad \Omega_{\gamma 0} = 2.47\times 
10^{-5}\, h_{0}^{-2}.
\label{rot11b}
\end{equation}
By taking the derivative of $\Omega_{\gamma}(x)$ with respect to $x$, the maximum of 
$\Omega_{\gamma}(x)$ is determined by the condition $ [4+ e^{x} ( x - 4)]=0$ 
which implies that $x_{\mathrm{max}} = 3.920$. 
The angular frequency related to $\nu_{\mathrm{max}}$  is given by 
 $\omega_{\max} = 2\pi \nu_{\mathrm{max}}$. By comparing 
 $\omega_{\mathrm{max}}$ with the other typical frequencies 
 of the problem we have that
\begin{equation}
 \omega_{\mathrm{max}} \gg \overline{\omega}_{\mathrm{pe}} \gg \overline{\omega}_{\mathrm{Be}} \gg 
\overline{\omega}_{\mathrm{pi}} \gg \overline{\omega}_{\mathrm{Bi}},
\label{EQ16}
\end{equation}
possible resonances in the dispersion relations are then avoided, but, still the 
polarization plane of the CMB can be Faraday rotated by an angle  
\begin{equation}
\Phi(\hat{n},\tau_{0}) = 
{\mathcal A}(\overline{\nu}) \int_{0}^{\tau_{0}} {\mathcal K}(\tau) \,[\hat{n} \cdot 
\vec{B}(\vec{x},\tau)]\,\, d\tau,\qquad {\mathcal K}(\tau) = \epsilon'(\tau) e^{- \epsilon(\tau,\tau_{0})}.
\label{rot1}
\end{equation}
In Eq. (\ref{rot1}) the squared brackets have been used to remind of the vector structure of the integrand; ${\mathcal A}(\overline{\nu}) = 3 \overline{\nu}^{-2}/(16 \pi^2 e)$ depends upon the comoving frequency $\overline{\nu}$; finally in Eq. (\ref{rot1})  ${\mathcal K}(\tau)$ is the visibility function
expressed in terms of the optical depth $\epsilon(\tau,\tau_{0})$
\begin{equation}
\epsilon(\tau,\tau_{0})  = \int_{\tau}^{\tau_{0}}\,\, x_{\mathrm{e}}(\tau')\, \,\sigma_{\mathrm{Th}} \,\,\tilde{n}_{\mathrm{e}}(\tau') a(\tau')\, d\tau'.
\label{rot2a}
\end{equation}
The visibility function vanishes  for  $\tau \gg \tau_{\mathrm{rec}}$
and has a maximum around recombination, i.e. when  
\begin{equation}
\epsilon'' + \epsilon'^2 =0, \qquad \frac{d \epsilon}{d \tau} = - x_{\mathrm{e}}(\tau) \tilde{n}_{\mathrm{e}}(\tau) \sigma_{\mathrm{Th}} a(\tau) \equiv - \epsilon'.
\label{rot2c}
\end{equation}
The second expression in Eq. (\ref{rot2c}) clarifies that a minus sign appears in the time derivative of $\epsilon(\tau,\tau_{0})$ since $\tau$ appears in the lower limit of integration.  
The magnetic fields not only enter the evolution of the charged 
species and of the photons (which are tightly coupled to electrons and baryons 
before decoupling) but also the (perturbed) Einstein equations.
The Hamiltonian and momentum constraints, stemming from the $(00)$ and $(0i)$ 
 components of the perturbed Einstein equations are:
\begin{eqnarray}
&& \nabla^2 \psi - 3 {\cal H} ( {\cal H}\phi + \psi') = 4\pi G a^2 [ \delta_{\mathrm{s}} \rho_{\rm t} + 
\delta_{\mathrm{s}}\rho_{\rm B}],\qquad \delta_{\mathrm{s}} \rho_{\rm B}(\tau,\vec{x}) = \frac{B^2(\vec{x})}{8\pi a^4(\tau)},
\label{Ham1}\\
&&\nabla^2( {\cal H} \phi + \psi') = - 4\pi G a^2 \biggl[(p_{\rm t} + \rho_{\rm t}) 
\theta_{\rm t} + \frac{\vec{\nabla} \cdot (\vec{E} \times\vec{B})}{4 \pi a^4}\biggr].
\label{Mom1}
\end{eqnarray}
In Eqs. (\ref{Ham1}) and (\ref{Mom1}) 
$\delta_{\mathrm{s}} \rho_{\rm t}$ and $\theta_{\rm t}$, denote, respectively, the total 
density fluctuation of the fluid mixture and the divergence of the total velocity field (i.e. $\theta_{\rm t}
= \partial_{i} v^{i}_{\rm t}$) whose expressions, in terms of the four 
components of the plasma, i.e. $\nu$, $\gamma$, ${\rm c}$ (CDM) and ${\rm b}$ (baryons), is 
\begin{equation}
\delta_{\mathrm{s}}\rho_{\rm t} = \sum_{{\rm a}} \delta_{\mathrm{s}}\rho_{\rm a},\qquad
\delta_{\mathrm{s}} p_{\rm t} = \sum_{{\rm a}} \delta_{\mathrm{s}} p_{\rm a},\qquad
(p_{\rm t} + \rho_{\rm t}) \theta_{\rm t} = \sum_{{\rm a}} (p_{\rm a} + \rho_{\rm a})
\theta_{\rm a}.
\label{sumdef}
\end{equation}
The spatial components of the perturbed Einstein equations, imply, instead 
\begin{eqnarray}
&& \psi'' + {\cal H} ( \phi' + 2 \psi') + ( 2 {\cal H}' + {\cal H}^2) \phi + 
\frac{1}{3} \nabla^2(\phi - \psi) = 4\pi G a^2 (\delta_{\mathrm{s}} p_{\rm t} + \delta_{\mathrm{s}} p_{\rm B}),
\label{tij}\\
&& \nabla^4 ( \phi - \psi) = 12 \pi G a^2 [ 
(p_{\nu} + \rho_{\nu}) \nabla^2 \sigma_{\nu} +
 (p_{\gamma} + \rho_{\gamma}) \nabla^2 \sigma_{\rm B}],\qquad 
 \delta_{\mathrm{s}} p_{\rm B} = \frac{\delta_{\mathrm{s}} \rho_{\rm B}}{3}.
 \label{anis1}
 \end{eqnarray}
In Eq. (\ref{anis1}) $\nabla^2 \sigma_{\nu}$ is the neutrino anisotropic stress, while  $\nabla^2 \sigma_{\rm B}$ is the magnetic field anisotropic 
stress defined as:
 \begin{equation}
 \nabla^2 \sigma_{\rm B} = 
 \frac{3}{16\pi a^4 \rho_{\gamma}} \vec{\nabla}\cdot [
  (\vec{\nabla}\times \vec{B})
 \times \vec{B}] + 
 \frac{\nabla^2 \Omega_{\rm B}}{4},\qquad \Omega_{\rm B}(\vec{x}) = \frac{\delta_{\mathrm{s}}\rho_{\rm B}(\tau, \vec{x})}{\rho_{\gamma}(\tau)},
 \label{magndef}
 \end{equation}
 where, $\Omega_{\rm B}(\vec{x})$ is the magnetic energy density referred to the photon energy density and it is constant to a very good approximation if the magnetic flux is frozen into the plasma element. 
 In closing this section it is appropriate to mention that, in the past, different 
 approaches to the effects of large-scale magnetic fields have been proposed.
In this respect the interested reader can usefully consult the the review articles reported in Refs. \cite{rev1}, \cite{rev2} and \cite{rev3}. 
\renewcommand{\theequation}{3.\arabic{equation}}
\setcounter{equation}{0}
\section{Autocorrelation of the Faraday rate}
\label{sec3}
The autocorrelation of the Faraday rotation rate can now 
be computed by taking into account the finite thickness 
of the last scattering surface.  This analysis will be performed 
both analytically and numerically.  The present results 
complement and extend the analysis of Ref.\cite{gk3} which differs 
from other attempts \cite{F0,F1,F2,F3,F4,F5,F6} insofar as the 
contribution of the magnetic fields is included at every step 
of the Einstein-Boltzmann hierarchy, from the 
initial conditions to the relevant dynamical equations. 

Faraday effect has been studied in the context of a uniform field approximation \cite{F0,F1,F2,F3,F4} which breaks explicitly spatial isotropy. As discussed in the previous section, it is not consistent to keep the magnetic field uniform and, still, to assign the angular power spectra as if spatial isotropy was unbroken. On the contrary
it has been correctly argued that the magnetic fields should not be uniform but rather stochastically distributed over different comoving scales as in Eqs. (\ref{EQ1})--(\ref{EQ2})  \cite{mg2,gk3,F5,F6}.
The physical situation is such that magnetic fields with comoving inhomogeneity scale of the order of the Mpc are uniform over the scale of the Larmor radius but are not uniform on the 
scale of the Hubble radius.  

In \cite{F5} the interesting observation has been that one could indeed compute, in the sudden decoupling, the correlation function of Faraday rotation.  A similar observation has been later made 
in \cite{F6} within the same approximations.  One limitation of \cite{gk3} is that, when computing the Faraday rotation angle, the visibility function has been taken to be infinitely thin  and the decoupling sudden. The latter approximation is known to be  inaccurate at 
small scales where the Faraday-induced B-mode autocorrelations increase.  
One of the themes of the present section will be to relax the sudden decoupling approximation in the calculation of the Faraday rate. 

\subsection{Sudden and smooth decoupling approximations}
The sudden decoupling approximation offers a reasonable estimate of the angular power 
spectrum of Faraday rotation but only for sufficiently large angular scales 
(i.e. for sufficiently low harmonics since $\theta \sim \pi/\ell$).  
Conversely the smooth decoupling approximation is relevant for large $\ell$ (in practice for $\ell \gg 150$) where the finite thickness of the last scattering surface (together with the other sources of diffusive damping)  determines the fall-off of the Faraday autocorrelations. 
In the sudden approximation the visibility function is sharply peaked around the recombination time with zero width, i.e. ${\mathcal K}(\tau) = \delta(\tau - \tau_{\mathrm{rec}})$.  
The rotation angle of Eq. (\ref{rot1}) can be expanded in ordinary spherical harmonics, i.e.   
 \begin{equation}
 \Phi(\hat{n},\tau_{0}) = \sum_{\ell m} f_{\ell m} Y_{\ell m}(\hat{n}), \qquad
 C_{\ell}^{(\mathrm{FF})} = \frac{1}{2\ell + 1} \sum_{m= -\ell}^{\ell} \langle f_{\ell m}^{*}\, f_{\ell m}\rangle, 
 \label{rot4a}
 \end{equation}
 where $C_{\ell}^{(\mathrm{FF})}$ denotes the autocorrelation of the Faraday rotation angle 
 in full analogy with the autocorrelations of the polarization observables 
 introduced in Eqs. (\ref{int4})--(\ref{int5}). By assuming that 
 ${\mathcal K}(\tau) = \delta(\tau - \tau_{\mathrm{rec}})$,
Eq. (\ref{rot1}) implies that
\begin{eqnarray}
\langle \Phi(\hat{n},\tau_{0}) \, \Phi(\hat{m},\tau_{0})\rangle &=& 
\sum_{\ell} \frac{2 \ell + 1}{ 4 \pi} C_{\ell}^{\mathrm{FF}} P_{\ell}(\hat{n}\cdot\hat{m}),
\label{rot3}\\
C_{\ell}^{(\mathrm{FF})} &=&  {\mathcal A}^2(\overline{\nu}) A_{\mathrm{B}} (k_{\mathrm{L}} \tau_{0})^{1 - n_{\mathrm{B}}} \, \pi^{3/2} \frac{\ell\,(\ell + 1)\,\Gamma\biggl(\frac{5}{2} - \frac{n_{\mathrm{B}}}{2}\biggr) \Gamma\biggl( \ell + \frac{n_{\mathrm{B}}}{2} - \frac{3}{2}\biggr)}{\Gamma\biggl(3 - \frac{n_{\mathrm{B}}}{2}\biggr) \Gamma\biggl( \frac{7}{2} + \ell - \frac{n_{\mathrm{B}}}{2} \biggr)},
\label{rot4}
\end{eqnarray}
Note that Eq. (\ref{rot4}) is derived under the approximation that 
$2(\ell + 1)  < 5 - n_{\mathrm{B}} < 0$ (see \cite{grad} p. 675).   
When the finite thickness effects of the last scattering surface 
are taken into account the visibility function can be approximated by a Gaussian profile centered at $\tau_{\mathrm{rec}}$, i.e. 
\begin{eqnarray}
{\mathcal K}(\tau) &=& {\mathcal N}(\sigma_{\mathrm{rec}}) e^{- \frac{(\tau - \tau_{\mathrm{rec}})^2}{2 \sigma_{\mathrm{rec}}^2}},\qquad \int_{0}^{\tau_{0}} {\mathcal K}(\tau) d\tau = 1, 
\label{rot5}\\
{\mathcal N}(\sigma_{\mathrm{rec}})&=& \sqrt{\frac{2}{\pi}} \frac{1}{\sigma_{\mathrm{rec}}} 
\biggl[ \mathrm{erf}\biggl( \frac{\tau_{0} - \tau_{\mathrm{rec}}}{\sqrt{2} \sigma_{\mathrm{rec}}} \biggr)+   \mathrm{erf}\biggl( \frac{ \tau_{\mathrm{rec}}}{\sqrt{2} \sigma_{\mathrm{rec}}} \biggr)\biggr]^{-1}, 
\label{rot6}\\
\mathrm{erf}(z) &=& \frac{2}{\sqrt{\pi}} \int_{0}^{z} e^{- t^2} dt.
\end{eqnarray}
The overall normalization ${\mathcal N}(\sigma_{\mathrm{rec}})$ has been 
chosen in such a way that the integral of ${\mathcal K}(\tau)$ is normalized to 
$1$: the visibility function is nothing but the probability that a photon last scatters between $\tau$ and $\tau + d\tau$. Equation
 (\ref{rot6}) simplifies when $\tau_{0} \gg \tau_{\mathrm{rec}}$ and 
 $ \tau_{0} \gg \sigma_{\mathrm{rec}}$, since, in this limit, the error 
 functions go  to a constant and ${\mathcal N}(\sigma_{\mathrm{rec}}) \to\sigma_{\mathrm{rec}}^{-1}\, \sqrt{2/\pi}$. In the latter limit, the thickness of the last scattering surface, i.e. $\sigma_{\mathrm{rec}}$, is of the order of $\tau_{\mathrm{rec}}$. In the explicit calculations the normalization constant is always taken in its full form (see Eq. (\ref{rot6})); however, to avoid lengthy expressions the final results 
will be expressed, when feasible, 
 in the limit $\tau_{0} \gg \tau_{\mathrm{rec}}$ and $\tau_{0} \gg \sigma_{\mathrm{rec}}$.  
 
Recalling that, in Eq. (\ref{rot1}) the magnetic field appears in real space, the $\Phi(\hat{n},\tau_{0})$ can also 
be written as: 
\begin{equation}
\Phi(\hat{n},\tau_{0}) = \frac{{\mathcal A}(\overline{\nu})}{( 2 \pi)^{3/2}} \int d^{3} k \,\, n^{i} B_{i}(\vec{k}) \int_{0}^{\tau_{0}} {\mathcal K}(\tau) \,e^{ - i k\mu(\tau_{0} - \tau)} \, d\tau,
\label{rot8}
\end{equation}
where the present position of the observer coincides with the 
origin and where $\mu = \hat{k} \cdot \hat{n}$ denotes, as usual, the projection 
of the momentum of the photon over the Fourier mode. Inserting Eq. (\ref{rot6}) 
into Eq. (\ref{rot8}) and performing the required integration over time 
the angular power spectrum becomes 
\begin{equation}
C_{\ell}^{(\mathrm{FF})} = 4 \pi {\mathcal A}^2(\overline{\nu}) \, 
\ell (\ell + 1) \int \frac{d k}{k} {\mathcal P}_{\mathrm{B}}(k) \frac{j_{\ell}^2(k\tau_{0})}{|k\tau_{0}|^2} \overline{{\mathcal K}}^2(k,\tau_{\mathrm{rec}}),\qquad \overline{{\mathcal K}}(k,\tau_{\mathrm{rec}}) = e^{- k^2/k_{\mathrm{t}}^2}, 
\label{rot9}
\end{equation}
where $j_{\ell}(z)$ are the spherical Bessel functions \cite{abr1}, i.e. 
\begin{equation}
j_{\ell}(z) = \sqrt{\frac{\pi}{2\,z}} J_{\ell + 1/2}(z).
\label{defsph}
\end{equation}
To derive Eq. (\ref{rot9}) it has been used that, thanks  to spatial isotropy \footnote{Recall that, by assumption, the stochastic background 
of relic magnetic fields does not break spatial isotropy and, therefore, $n^{i} n^{j} = \delta^{ij}/3$.}, 
$(\hat{k}\cdot\hat{n})^2 = k^2/3$
 and where we defined $k_{\mathrm{t}}= \sqrt{6}/\sigma_{\mathrm{rec}}$.
The result of Eq. (\ref{rot9}) follows directly from Eq. (\ref{rot4a}). More specifically, it can be shown that the $f_{\ell m}$ are 
determined to be 
\begin{equation}
f_{\ell m} = \frac{4\pi (-i)^{\ell -1}}{(2\pi)^{3/2}} {\mathcal A}(\overline{\nu}) \sqrt{\ell (\ell + 1)} \int d^{3} k \frac{j_{\ell} (k\tau_{0})}{|k\tau_{0}|} \vec{B}(\vec{k})\cdot \vec{Y}^{(1)*}_{\ell m}(\hat{k})  \overline{{\mathcal K}}(k,\tau_{\mathrm{rec}}),
\label{rot9a}
\end{equation}
where $\vec{Y}^{(1)}_{\ell m}(\hat{k}) = \vec{\nabla}_{\hat{k}} Y_{\ell m}(\hat{k})/\sqrt{\ell (\ell + 1)}$ is the vector harmonic
which arises from the analog of the Rayleigh expansion but in the case of a (solenoidal) vector field 
\cite{bie,var} (like the magnetic field). Inserting Eq. (\ref{rot9a}) inside Eq. (\ref{rot4a}), Eq. (\ref{rot9}) is swiftly recovered. The angular power spectrum can then be written, in a compact notation, as 
\begin{eqnarray}
&&C_{\ell}^{(\mathrm{FF})} = \overline{C}^{(\mathrm{FF})} \,\, \ell(\ell + 1)\, \, {\mathcal I}_{\ell}(n_{\mathrm{B}}, k_{\sigma}, k_{\mathrm{D}}, k_{\mathrm{t}}),
\label{rot10}\\
&&\overline{C}^{(\mathrm{FF})}  = 4\pi {\mathcal A}^2(\overline{\nu}) A_{\mathrm{B}} \biggl(\frac{k_{0}}{k_{\mathrm{L}}}\biggr)^{n_{\mathrm{B}}-1} \equiv 1.015\times 10^{-5}\,\,\bigg(\frac{A_{\mathrm{B}}}{\mathrm{nG}^2}\biggr) \biggl(\frac{\nu}{\nu_{\mathrm{max}}}\biggr)^{-4}\,
\biggl(\frac{k_{0}}{k_{\mathrm{L}}}\biggr)^{n_{\mathrm{B}} -1}.
\label{rot10a}
\end{eqnarray}
The integral appearing in Eq. (\ref{rot10}) is defined as 
\begin{equation}
{\mathcal I}_{\ell}(n_{\mathrm{B}}, k_{\sigma}, k_{\mathrm{D}}, k_{\mathrm{t}}) = 
\int_{0}^{\infty} x_{0}^{n_{\mathrm{B}} -4} j_{\ell}^2(x_{0}) e^{ - 2 (x_{0}/x_{\mathrm{d}})^2}\,\, d x_{0},
\label{rot11}
\end{equation}
where, on top of $k_{\mathrm{t}}$ (already encountered in Eq. (\ref{rot9}))
$k_{\mathrm{D}}$ and $k_{\sigma}$ parametrize, respectively, 
the effects of the thermal and magnetic diffusivities \cite{alfven,krall}, i.e. 
$x_{\mathrm{d}} = k_{\mathrm{d}}\tau_{0}$ where 
\begin{equation}
k_{\mathrm{d}}^{-2} = k_{\mathrm{D}}^{-2} + k_{\sigma}^{-2} + k_{\mathrm{t}}^{-2}.
 \label{rot12}
 \end{equation}
Note that, in Eq. (\ref{rot10a}), the normalization is fixed by choosing, as pivot frequency, $\nu_{\mathrm{max}}$ 
which maximizes the spectral energy density of the CMB (see Eqs. (\ref{rot11a})--(\ref{rot11b}) and discussion therein). 
The three separate contributions arising in Eq. (\ref{rot12}) will now be estimated. 
 
As discussed in Eq. (\ref{rot2c}) the maximum 
of the visibility function is determined by the condition  $\epsilon'' + \epsilon'^2 \simeq 0$
which implies, in terms of the ionization fraction, that 
\begin{equation}
x_{\mathrm{e}}' - 2 {\mathcal H} x_{\mathrm{e}} + \tilde{n}_{\mathrm{e}} \sigma_{\mathrm{Th}} x_{\mathrm{e}}^2 \simeq 0.
\label{rot14a}
\end{equation}
Equation (\ref{rot14a}) defines $\tau_{\mathrm{rec}}$, indeed, at $\tau_{\mathrm{rec}}$, 
$x_{\mathrm{rec}}' \simeq - \sigma_{\mathrm{Th}} a(\tau_{\mathrm{rec}}) \tilde{n}_{\mathrm{e}}(\tau_{\mathrm{rec}})$. The ionization fraction can be independently obtained from the approximate solution 
of the equation for recombination \cite{zeld1}
\begin{equation}
x_{\mathrm{e}}(z) = \frac{1.086\times 10^{8}}{z} 
\biggl(\frac{h_{0}^2 \, \Omega_{\mathrm{M}0}}{0.134}\biggr)^{1/2}\, \biggl(\frac{h_{0}^2 \Omega_{\mathrm{b}0}}{0.023}\biggr)^{-1} \,e^{- B/z},\qquad B \simeq 1.3\times 10^{4}.
\label{rot14b}
\end{equation}
A slightly different parametrization can be found in Ref. \cite{wyse}.
From Eqs. (\ref{rot14a})--(\ref{rot14b}) it follows that the visibility 
function peaks, within the present approximations, for $z_{\mathrm{rec}} \simeq 1050$ 
and that $x^{\prime}_{\mathrm{e}}(\tau_{\mathrm{rec}}) \simeq {\mathcal H}_{\mathrm{rec}} (B/z_{\mathrm{rec}}) \sigma_{\mathrm{Th}} \tilde{n}_{\mathrm{e}}(\tau_{\mathrm{rec}}) a(\tau_{\mathrm{rec}})$. 
Introducing therefore the parameter $\alpha_{\mathrm{t}} \simeq \sqrt{3} B/z_{\mathrm{rec}}$ we will have that $\ell_{\mathrm{t}} = k_{\mathrm{t}} \tau_{0}$ can be written as
\begin{equation}
\ell_{\mathrm{t}}= 
\frac{2\alpha_{\mathrm{t}} \sqrt{z_{\mathrm{rec}}}}{
{\mathcal I}_{\Lambda}} 
\sqrt{1 + \frac{z_{\mathrm{eq}}}{z_{\mathrm{rec}}}},
\label{rot15a}
\end{equation}
where 
\begin{equation}
{\mathcal I}_{\Lambda} = 3\biggl(
\frac{\Omega_{\Lambda 0}}{\Omega_{\mathrm{M}0}}\biggr)^{1/6}  I_{\Lambda}^{-1},\qquad
I_{\Lambda} = \int_{0}^{\mathrm{arcsinh}{\sqrt{(\Omega_{\Lambda 0}/\Omega_{\mathrm{M}0})}}} \frac{dy}{\sinh^{2/3}{y}}.
\label{rot15aa}
\end{equation}
Since we will always be assuming the flat $\Lambda$CDM paradigm, Eq. (\ref{rot15aa})
depends effectively only upon $\Omega_{\mathrm{M}0}$. The numerical result for the integral 
can be written as
${\mathcal I}_{\Lambda} = \Omega_{\mathrm{M}0}^{- \gamma_{1}}$ and  $\gamma_{1} = - 0.0858$.

The shear viscosity term (to a given order in the tight-coupling expansion)
allows for the estimate of $k_{\mathrm{D}}$. 
To zeroth order in the tight-coupling expansion the monopole and the dipole of the radiation field obey, respectively, the following pair of equations
\begin{eqnarray}
&& \Delta_{\mathrm{I}0}'' + \frac{{\mathcal   H} R_{\rm b}}{R_{\rm b} + 1} \Delta_{\mathrm{I}0}' + 
\frac{4}{15} k^2 \frac{\lambda_{\mathrm{Th}}}{R_{\mathrm{b}} +1} \Delta_{\mathrm{I}0}'+
\frac{k^2}{3 ( R_{\rm b} + 1)} \Delta_{\mathrm{I}0} =  \biggl[ \psi'' + 
\frac{{\mathcal H} R_{\rm b}}{R_{\rm b} + 1} \psi' - \frac{k^2}{3} \phi\biggr] 
\nonumber\\
&& + 
\frac{k^2 }{3 ( R_{\rm b} + 1)} \biggl[  \sigma_{\rm B} - \frac{\Omega_{\rm B}}{4}\biggr],
\label{MON}\\
&& \Delta_{\mathrm{I}1}' + \frac{{\mathcal H} R_{\mathrm{b}}}{1 + R_{\mathrm{b}}}\Delta_{\mathrm{I}1} + \frac{4 \lambda_{\mathrm{Th}}}{15(R_{\mathrm{b}} + 1)} k \Delta_{\mathrm{I}1}= 
\frac{k}{3 ( 1 + R_{\mathrm{b}})} \Delta_{\mathrm{I}0} + \frac{k}{3}\phi + \frac{k (\Omega_{\mathrm{B}} - 4 \sigma_{\mathrm{B}})}{12 ( 1 + R_{\mathrm{b}})},
\label{DIP}
\end{eqnarray}
where $\lambda_{\mathrm{Th}} = 1/\epsilon'$. In the absence of $\lambda_{\mathrm{Th}}$, Eqs. (\ref{MON}) 
and (\ref{DIP}) can be derived from Eqs. (\ref{DC1}) and (\ref{EQ10}) by recalling that, by definition, 
the monopole and the dipole of the intensity of the radiation field are related, respectively, to the 
density contrast and to baryon-photon velocity, i.e. $\delta_{\gamma} = 4 \Delta_{\mathrm{I}0}$ 
and  $\theta_{\gamma\mathrm{b}} = 3\,k\, \Delta_{\mathrm{I}1}$.
Introducing the photon-baryon sound speed 
\begin{equation}
c_{\mathrm{sb}}^2(\tau) = \frac{1}{3 [ 1 + R_{\mathrm{b}}(\tau)]},\qquad R_{\mathrm{b}}(z) = \frac{3}{4} \frac{\rho_{\mathrm{b}}}{\rho_{\gamma}}  = 0.664 \biggl(\frac{ h_{0}^2 \Omega_{\mathrm{b}0}}{0.023}\biggr) \biggl(\frac{1051}{z_{\mathrm{rec}} + 1}\biggr),
\label{rot12b}
\end{equation}
in Eq. (\ref{MON}) the variable can be changed from $\tau$ to $d q = c_{\mathrm{sb}}^2 d\tau$. The 
resulting equation is solvable the WKB approximation \cite{mg3}. 
Since we are dealing here with the polarization observables 
we are specifically interested in the evolution of the dipole. In the most simplistic case, i.e. when only the adiabatic mode is present
\begin{equation}
\Delta_{\mathrm{I}1}(k,\tau) = - \frac{\sqrt{c_{\mathrm{sb}}}\,{\mathcal R}_{*}(k)}{5} \sin{[k\, r_{\mathrm{s}}(\tau)]} e^{- k^2/k_{\mathrm{D}}^2},\qquad r_{\mathrm{s}}(\tau) = \int_{0}^{\tau} \, d\tau' \, c_{\mathrm{sb}}(\tau'),
\label{SP7}
\end{equation}
where ${\mathcal R}_{*}(k)$ is a stochastic variable obeying Eq. (\ref{EQ2a}) and where 
$k_{\mathrm{D}}$ can be estimated as:
\begin{equation}
\frac{1}{k^2_{\mathrm{D}}} = \frac{2}{5} \int_{0}^{\tau} c_{\mathrm{sb}}(\tau') \frac{  d\tau'}{a(\tau')\,\,x_{\mathrm{e}} n_{\mathrm{e}} \sigma_{\mathrm{Th}}}.
\label{rot12a}
\end{equation}
The estimate of Eq. (\ref{rot12a}) can be further improved by going 
to second order in the tight-coupling expansion where the inclusion of the polarization fluctuations allows for a 
slightly more accurate estimate \cite{zal}:
\begin{equation}
\frac{1}{k_{\mathrm{D}}^2 }= \int_{0}^{\tau} \frac{d\tau'}{6 (R_{\mathrm{b}} + 1) \epsilon'} \biggl[\frac{16}{15} + \frac{R_{\mathrm{b}}^2}{R_{\mathrm{b}} + 1}\biggr].
\label{rot13}
\end{equation}
The  factor $16/15$ arises since the polarization fluctuations are taken consistently 
into account in the derivation. 
To obtain explicit estimates of the diffusive damping it is very practical to use an explicit form 
of the scale factor. It can be shown by direct substitution that, across the matter radiation 
equality the solution of Eqs. (\ref{FL1})--(\ref{FL2}) 
is given by $a(\tau) = a_{\mathrm{eq}} [ (\tau/\tau_{1})^2 + 2(\tau/\tau_{1})]$.  To zeroth-order in the tight-coupling expansion, after a lengthy but straightforward calculation, we will have 
\begin{equation} 
\frac{1}{ \overline{k}_{\mathrm{D}} \tau_{\mathrm{rec}}} = 9.63\times 10^{-3} \biggl(\frac{h_{0}^2\Omega_{\mathrm{b}0}}{0.023}\biggr)^{-1/2}
\biggl(\frac{h_{0}^2\Omega_{\mathrm{M}0}}{0.134}\biggr)^{1/4} \biggl(\frac{1051}{z_{\mathrm{rec}}}\biggr)^{3/4}.
\label{rot13a}
\end{equation}
This estimate  can be obtained from Eq. (\ref{rot12a}) by setting $R_{\mathrm{b}} =0$ in the
baryon-photon sound speed. More realistically we have instead that 
\begin{equation}
R_{\mathrm{b}}(\tau) = \overline{R}_{\mathrm{b}} \biggl[ \biggr(\frac{\tau}{\tau_{1}}\biggl)^2 + 2\frac{\tau}{\tau_{1}}\biggr],\qquad 
\overline{R}_{\mathrm{b}} = 0.216 \biggl(\frac{h_{0}^2 \Omega_{\mathrm{b}0}}{0.023}\biggr) \biggl(\frac{h_{0}^2 \Omega_{\mathrm{M}0}}{0.134}\biggr)^{-1}.
\label{rot13b}
\end{equation}
In this case the estimate changes quantitatively but it can still be expressed as in Eq. (\ref{rot13a}) but with a different 
pre-factor, i.e. $ 9.63\times 10^{-3}$ becomes $2.17\times 10^{-2}$. More accurate expressions 
can be obtained from the second-order treatment by performing numerically the various integrations and by using, repeatedly, the interpolating form of the scale factor. 

The further source of damping contemplated by Eq. (\ref{rot12}) is the one connected with the finite value of 
the conductivity. In the electron-ion plasma characterizing our physical system the conductivity is determined by Coulomb scattering, i.e. 
\begin{equation}
\sigma_{\mathrm{Coul}} = \frac{\alpha_{\mathrm{em}}^2}{T^2} \ln{\Lambda_{\mathrm{C}}},
\qquad \Lambda_{\mathrm{C}} = \frac{3}{2} \biggl(\frac{T^3}{\pi n_{\mathrm{e}}}\biggr)^{1/2} \frac{1}{e^3}.
\label{rot13c}
\end{equation}
To compute the conductivity we can assume that each collision produces a 
large deflection. Under this approximation we will have that 
\begin{equation}
\overline{\sigma} = \frac{\omega_{\mathrm{pe}}^2}{4\,\pi\,\Gamma_{\mathrm{Coul}}}
= \sigma_{\mathrm{c}} \, a, \qquad
\sigma_{\mathrm{c}} = \frac{T}{\alpha_{\mathrm{em}}\ln{\Lambda_{\mathrm{c}}}} \biggl(\frac{T}{m_{\mathrm{e}}}\biggr)^{1/2},
\label{rot13d}
\end{equation}
where
$\Gamma_{\mathrm{Coul}} = \tilde{n}_{\mathrm{e}} \, x_{\mathrm{e}}\,v_{\mathrm{th}} \sigma_{\mathrm{Coul}}$ is the Coulomb rate.
The typical conductivity scale $k_{\sigma}^{-1}$ is then determined according to
\begin{equation}
\frac{1}{k_{\sigma}^2} =  \frac{\tau_{\mathrm{rec}}}{4\pi \overline{\sigma}},
\label{rot14}
\end{equation}
and it is typically smaller than the diffusive damping.
The Silk damping is often introduced by summing up the two dissipative contributions, i.e. 
 by defining $\ell_{\mathrm{S}}^{-2} = \ell_{\mathrm{t}}^{-2} + \ell_{\mathrm{D}}^{-2}$ which also 
 implies that 
 \begin{equation}
 \ell_{\mathrm{S}} = \frac{\ell_{\mathrm{t}}}{\sqrt{(k_{\mathrm{t}}/k_{\mathrm{D}})^2 +1}}.
\label{rot15b}
 \end{equation}
 
\subsection{Explicit forms of the Faraday autocorrelations at small angular scales}

The integral of Eq. (\ref{rot11}) can be evaluated in a closed form in terms of generalized hypergeometric 
functions \cite{grad}:
 \begin{equation}
{\mathcal I}_{\ell}(n_{\mathrm{B}}, k_{\mathrm{d}}) = 2^{- 3\ell -5/2}\, (k_{\mathrm{d}} \tau_{0})^{2\ell + n_{\mathrm{B}} - 3} 
\frac{\Gamma\biggl(\frac{n_{\mathrm{B}}}{2} - \frac{3}{2} + \ell\biggr)}{\Gamma^2\biggl(\ell +\frac{3}{2}\biggr)}\,
F_{3 3}\biggl[ a_{1},\,a_{2},\,a_{3};\, b_{1},\,b_{2},\,b_{3}; - \frac{k_{\mathrm{d}}^2\tau_{0}^2}{2}\biggr],
\label{rot16}
\end{equation}
where, by definition of generalized hypergeometric, two of the three arguments are replaced by 
generalized vectors of dimensions $q$ and $p$, 
i.e. $F_{q p}(\vec{a}; \vec{b}; z) \equiv F_{q p}(a_{1},...,a_{p};\, b_{1},...,b_{q}; z)$. In the case 
$q = p = 3$ and for the case of Eq. (\ref{rot16}) the components of the two generalized vectors are given by:
\begin{eqnarray}
&& a_{1} = \ell + 1,\quad a_{2} = \ell + \frac{3}{2},\qquad a_{3} = \ell + \frac{n_{\mathrm{B}}}{2} - \frac{3}{2},
\label{rot16a}\\
&& b_{1} = \ell + \frac{3}{2},\qquad b_{2} = \ell + \frac{3}{2},\qquad b_{3} = 2 \ell + 2.
\label{rot16b}
\end{eqnarray}
For the present purposes Eq. (\ref{rot16}) is not so revealing. More explicit expressions can be 
obtained by working out the large $\ell$ limit of the spherical functions 
$j_{\ell}(x_{0})$. 
In the limit $\ell \gg 1$ the spherical Bessel functions can be expressed as \cite{abr1,abr2}
\begin{equation}
j_{\ell}^2( \ell w) \simeq \frac{\cos^2{[\beta(w,\ell)]}}{\ell w \sqrt{ \ell^2 w^2 - \ell^2}}, \qquad \beta(w,\ell) = \ell \sqrt{w^2 -1} - \ell\arccos{(w^{-1})} - \frac{\pi}{4}, 
\label{rot17}
\end{equation}
where $w > 1$ and $x_{0} = \ell \, w$. Using Eq. (\ref{rot17}) inside Eq. (\ref{rot11}) it is easy to obtain:
\begin{equation}
{\mathcal I}_{\ell}(n_{\mathrm{B}}, \ell_{\mathrm{d}}) = \ell^{n_{\mathrm{B}} -5} \int_{1}^{\infty} 
\frac{\cos^2{[\beta(w,\ell)]}}{\sqrt{w^2 -1}} \, w^{n_{\mathrm{B}} -5}\, e^{- 2 (\ell/\ell_{\mathrm{d}})^2 w^2} \, dw.
\label{rot17a}
\end{equation}
Changing integration variable as $w = \sqrt{y^2 +1}$  we do get that
\begin{equation}
{\mathcal I}_{\ell}(n_{\mathrm{B}}, \ell_{\mathrm{d}}) = \ell^{n_{\mathrm{B}}-5} 
\int_{0}^{\infty}\,\, dy\,\, (y^2 + 1)^{(n_{\mathrm{B}} - 6)/2} e^{- 2\frac{\ell^2}{\ell_{\mathrm{d}}^2} (y^2 + 1)},
\label{rot18}
\end{equation}
where the oscillatory contribution is, in practice, averaged so that it is fully legitimate, in the large $\ell$ limit, to 
demand $\cos^2[\beta(w,\ell)]\to 1/2$. 
The integral over $y$ arising in Eq. (\ref{rot18}) can be performed in explicit terms and it can be always expressed as
\begin{equation}
{\mathcal I}_{\ell}(n_{\mathrm{B}}) = \frac{\ell^{n_{\mathrm{B}} -5}}{2} \,\, {\mathcal G}(z, n_{\mathrm{B}})\,\,\, e^{-2 \frac{\ell^2}{\ell_{\mathrm{d}}^2}},\qquad z = \frac{\ell}{\ell_{\mathrm{d}}}  ,
\label{rot19}
\end{equation}
where 
\begin{equation}
{\mathcal G}(z,n_{\mathrm{B}}) =\sqrt{\pi} \frac{\Gamma\biggl(\frac{5- n_{\mathrm{B}}}{2} \biggr)}{\Gamma\biggl(\frac{6- n_{\mathrm{B}}}{2}\biggr)} 
F_{11}\biggl[\frac{1}{2}, \frac{n_{\mathrm{B}}-3}{2}, 2 z^2\biggr] + (2\,z)^{(5-n_{\mathrm{B}})/2}
\frac{ \Gamma\biggl(\frac{n_{\mathrm{B}}- 5}{2}\biggr)}{\Gamma\biggl(\frac{6- n_{\mathrm{B}}}{2}\biggr)} 
F_{11}\biggl[ \frac{6 - n_{\mathrm{B}}}{2}, \frac{7- n_{\mathrm{B}}}{2}, 2 z^2\biggr].
\label{rot20}
\end{equation}
In the last equation, $F_{11}(c,d, z)$ denotes the confluent hypergeometric function \cite{abr1,abr2}. As 
a function of $z$, ${\mathcal G}(z,n_{\mathrm{B}})$ is  slowly varying and it is always 
of order $1$ for a variation of $z$ between, say $0.001$ and, e.g. $10$ which is the physical range of harmonics. 
The result expressed by Eq. (\ref{rot20}) seems to be naively singular when $n_{\mathrm{B}}$ is a natural number 
since, in this case, the Euler Gamma function has a negative (integer) argument. This divergence is just apparent since by going back 
to the integral of Eq. (\ref{rot18}) it can be explicitly checked that the result is convergent (both in the origin and at 
infinity) also when $n_{\mathrm{B}}$ is a natural number. As an example consider the case $n_{\mathrm{B}} =1$; in this 
case the result of the direct integration appearing in Eq. (\ref{rot18}) can always be recast in the form 
of Eq. (\ref{rot19}) but where 
\begin{equation}
{\mathcal G}(z,1) = \frac{2}{3} z^2 e^{z^2}[ 2 z^2 K_{0}(z^2) + (1 - 2 z^2) K_{1}(z^2)],
\label{rot20a}
\end{equation}
where $K_{0}(z)$ and $K_{1}(z)$ are  analytically continued Bessel functions. In spite 
of the presence of the (increasing) exponential factor, ${\mathcal G}(z,1)$ 
approaches zero for large $z$. For $z\to 0$, ${\mathcal G}(z,1)\to 2/3$.  It is amusing to notice that this 
result is consistent with Eq. (\ref{rot20}). Indeed, in the limit $z\to 0$,  ${\mathcal G}(z, n_{\mathrm{B}}) =
 \sqrt{\pi}/2[\Gamma((5-n_{\mathrm{B}})/2]/\Gamma(3- n_{\mathrm{B}}/2)$ which equals $2/3$ for $n_{\mathrm{B}} = 1$. 

\renewcommand{\theequation}{4.\arabic{equation}}
\setcounter{equation}{0}
\section{Semi-analytic estimates of the polarization observables}
\label{sec4}
As discussed in the introduction, the calculation of the B-mode polarization 
is done, for our semi-analytic purposes, by iteration.  In the first part of the present 
section, by applying previous methods described in 
\cite{mg2,mg3}, we ought to derive a semi-analytic expression for the E-mode autocorrelation which 
could be extrapolated for large multipoles with reasonable accuracy.  In the second part 
of this section, the results of Section \ref{sec3} will be combined with 
the estimates of the E-mode autocorrelations and we shall be able of obtaining 
a closed expression, valid at small angular scales, for the B-mode autocorrelations. 

\subsection{E-mode autocorrelations}
From the known expressions of the polarization observables (see, e.g.  Eqs. (\ref{int1})--(\ref{int4})) 
it is easy to see directly that the presence of a B-mode polarization is directly  related to a mismatch 
between $\Delta_{+}(\hat{n},\tau)$ and $\Delta_{-}(\hat{n},\tau)$. This statement is clear in real space where 
\begin{eqnarray}
&& \Delta_{\mathrm{E}}(\hat{n},\tau) = - \frac{1}{2} \{ K_{-}^{(1)}(\hat{n})[K_{-}^{(2)}(\hat{n})
\Delta_{+}(\hat{n},\tau)] +  K_{+}^{(-1)}(\hat{n})[K_{+}^{(-2)}(\hat{n}) \Delta_{-}(\hat{n},\tau)]\},
\label{BE1}\\
&&  \Delta_{\mathrm{B}}(\hat{n},\tau) =  \frac{i}{2} \{ K_{-}^{(1)}(\hat{n})[K_{-}^{(2)}(\hat{n})
\Delta_{+}(\hat{n},\tau)] -  K_{+}^{(-1)}(\hat{n})[K_{+}^{(-2)}(\hat{n}) \Delta_{-}(\hat{n},\tau)]\}.
\label{BE2}
\end{eqnarray}
The differential operators appearing in Eqs. (\ref{BE1}) and (\ref{BE2}) 
are  generalized ladder operators \cite{sud,zalda} whose action 
either raises or lowers the spin weight of a given fluctuation. They are defined, in general terms, as 
acting on a fluctuation of spin weight $s$: 
\begin{eqnarray}
&& K_{+}^{\mathrm{s}}(\hat{n}) = - (\sin{\vartheta})^{\mathrm{s}}\biggl[ \partial_{\vartheta} + 
\frac{i}{\sin{\vartheta}} \partial_{\varphi}\biggr] \frac{1}{(\sin{\vartheta})^{\mathrm{s}}},
\label{Kp}\\
&& K_{-}^{\mathrm{s}}(\hat{n}) = - \frac{1}{(\sin{\vartheta})^{\mathrm{s}}}
\biggl[ \partial_{\vartheta} -
\frac{i}{\sin{\vartheta}} \partial_{\varphi}\biggr] (\sin{\vartheta})^{\mathrm{s}}.
\label{Km}
\end{eqnarray}
For instance, $K_{-}^{(2)}\Delta_{+}$ will transform, overall, as a function of spin-weight 1 while 
$K_{-}^{(1)}[K_{-}^{(2)}\Delta_{+}]$ is, as anticipated, as scalar. Using Eqs. (\ref{Kp}) and (\ref{Km}) 
inside Eqs. (\ref{BE1}) and (\ref{BE2}) the explicit expressions of the E-mode and of the B-mode 
are, in real space:
\begin{eqnarray}
\Delta_{\mathrm{E}} (\hat{n},\tau) &=& - \frac{1}{2}\biggl\{( 1 -\mu^2) \partial_{\mu}^2 (\Delta_{+} + \Delta_{-}) - 4 \mu  \partial_{\mu}(\Delta_{+} + \Delta_{-}) - 2  (\Delta_{+} + \Delta_{-}) 
\nonumber\\
&-& 
\frac{\partial_{\varphi}^2 (\Delta_{+} + \Delta_{-})}{1 - \mu^2 } 
+
2 i\biggl[ \partial_{\varphi}  \partial_{\mu}(\Delta_{+} - \Delta_{-}) - \frac{\mu}{1 - \mu^2} \partial_{\varphi}  (\Delta_{+} - \Delta_{-})\biggr] \biggr\},
\label{BE3}\\
 \Delta_{\mathrm{B}} (\hat{n},\tau) &=& \frac{i}{2} \biggl\{( 1 -\mu^2) \partial_{\mu}^2(\Delta_{+} - \Delta_{-}) - 4 \mu  \partial_{\mu}(\Delta_{+} - \Delta_{-}) - 2  (\Delta_{+} - \Delta_{-})  
\nonumber\\
&-& 
\frac{\partial_{\varphi}^2  (\Delta_{+} - \Delta_{-})}{1 - \mu^2 }
+ 
2 i\biggl[ \partial_{\varphi}  \partial_{\mu}(\Delta_{+} + \Delta_{-}) - \frac{\mu}{1 - \mu^2} \partial_{\varphi}  (\Delta_{+} + \Delta_{-})\biggr] \biggr\}.
\label{BE4}
\end{eqnarray}
In Eqs. (\ref{BE3}) and (\ref{BE4}) $\partial_{\mu}$ denotes a derivation 
with respect to $\mu = \cos{\vartheta}$ while $\partial_{\varphi}$ denotes a derivation 
with respect to $\varphi$. Equations (\ref{BE3}) and (\ref{BE4}) hold in the spherical basis 
where $\hat{n}= (\vartheta,\varphi)$.
From Eqs. (\ref{BE3}) and (\ref{BE4}) it can be concluded that the B-mode polarization is absent 
(i.e. $\Delta_{\mathrm{B}}( \hat{n},\tau) =0$) 
provided  $\Delta_{+}(\hat{n},\tau)= \Delta_{-}(\hat{n},\tau)$ 
and provided $\partial_{\varphi}( \Delta_{+} + \Delta_{-}) =0$. The latter 
condition implies that the brightness perturbation is independent upon $\varphi$. Indeed,
to lowest order in our iterative scheme we will have that 
$\Delta_{+}(\hat{n},\tau) = \Delta_{-}(\hat{n},\tau) = \Delta_{\mathrm{P}}(\hat{n},\tau)$. Moreover, in Fourier space   
$\Delta_{\mathrm{P}}(k,\mu,\tau)$ does not depend upon $\varphi$ and it obeys 
\begin{equation}
\Delta_{\mathrm{P}}' + (i k \mu + \epsilon') \Delta_{\mathrm{P}} =  \frac{3 \, \epsilon'}{4} (1 -\mu^2) S_{\mathrm{P}}, \qquad S_{\mathrm{P}} = (\Delta_{\mathrm{I}2} + \Delta_{\mathrm{P}0} + \Delta_{\mathrm{P}2}),
\label{BE6}
\end{equation}
where $S_{\mathrm{P}}(k,\tau)$ 
is computed consistently 
with the other evolution equations which include the 
large-scale magnetic fields both at the level of the initial conditions as well as at the level
of the dynamical equations. 
To lowest order, the heat transfer equation does 
not depend upon $\varphi$ the following pair of equations will hold, in real space:
\begin{equation}
\Delta_{\mathrm{E}}(\hat{n},\tau) = - \partial_{\mu}^2[(1 - \mu^2) \Delta_{\mathrm{P}}(\hat{n},\tau)], \qquad 
a_{2,\,\,\ell\,m} = a_{-2,\,\,\ell\,m}.
\label{BE7}
\end{equation}
As a consequence of the second relation in Eq. (\ref{BE7}), it also follows 
that 
\begin{eqnarray}
a_{\ell\, m}^{(\mathrm{E})} &=& - \frac{1}{2} \int d \hat{n} \Delta_{\mathrm{P}}(\vec{x},\hat{n},\tau_{0}) [ _{2} Y_{\ell\, m}^{*}(\hat{n}) + _{-2} Y_{\ell\, m}^{*}(\hat{n})],
\label{BE8}\\
_{\pm 2}Y_{\ell\,m}(\hat{n}) &=& N_{\ell} \biggl[\partial_{\vartheta}^2 - \cot{\vartheta}\, \partial_{\vartheta} \pm 
\frac{2 i}{\sin{\vartheta}}(\partial_{\vartheta} - \cot{\vartheta})\partial_{\varphi} - \frac{1}{\sin^2{\vartheta}} \partial_{\varphi}^2\biggr] Y_{\ell\,m}(\hat{n}),
\label{BE9}
\end{eqnarray}
where $_{\pm}Y_{\ell\,m}(\hat{n})$ are the celebrated spin-2 spherical harmonics. 
Using line-of-sight integration, the autocorrelation of the E-mode polarization (defined in Eq. (\ref{int5})) can be written as
\begin{eqnarray}
C_{\ell}^{(\mathrm{EE})} &=& 4\pi \int \frac{d k}{k} {\mathcal P}_{{\mathcal R}}(k) |\Delta^{(\mathrm{E})}_{\ell}(k)|^2 
\label{CLEE1}\\
\Delta^{(\mathrm{E})}_{\ell}(k) &=&\frac{3}{4} \sqrt{\frac{(\ell + 2)!}{(\ell -2)!}} \int_{0}^{\tau_{0}} {\mathcal K}(\tau) S_{\mathrm{P}}
\frac{j_{\ell}(x_0)}{x_{0}^2}.
\label{CLEE2}
\end{eqnarray}
The E-mode autocorrelations of Eq.  (\ref{CLEE2}) can be estimated 
with semi-analytic methods.  For this purpose the first step is 
to derive $S_{\mathrm{P}}(k,\tau)$ in the 
tight-coupling approximation. This first step  is known to be insufficient for a reliable estimate of the E-mode autocorrelation. 
Therefore, following 
the approach originally developed in \cite{zal}, the tight-coupling 
expansion can be systematically 
improved by deriving an effective evolution equation for 
$S_{\mathrm{P}}$, i.e. 
\begin{equation}
 S_{\rm P}' + \frac{3}{10} \epsilon' S_{\rm P} = k \frac{2}{5} \overline{\Delta}_{{\rm I}1}(k,\tau),
\label{SP5}
 \end{equation}
 where $\overline{\Delta}_{\mathrm{I} 1}(k,\tau)$ is the dipole of the intensity (related to the velocity 
 of the photons) and evaluated to zeroth-order in the tight-coupling expansion. Using the solution of Eq. (\ref{SP5}) we then have 
  \begin{equation}
 \int_{0}^{\tau_{0}}d\tau {\mathcal K}(\tau) S_{\mathrm{P}}(k,\tau) = - \frac{2}{5} k \overline{\Delta}_{\mathrm{I}1} \sigma_{\mathrm{rec}} \int_{0}^{\infty} d\epsilon \, e^{-\frac{7}{10}\epsilon} \int_{1}^{\infty} e^{- 3\epsilon\,y/10} \frac{d y}{y}.
 \label{SP6}
 \end{equation}
Recalling Eq. (\ref{SP7})
we can express the angular power spectrum of the E-mode polarization as
\begin{eqnarray}
C_{\ell}^{(\mathrm{EE})}&=& \overline{C}^{(\mathrm{EE})} \ell (\ell + 1) (\ell + 2) (\ell -1)\biggl(\frac{k_{0}}{k_{\mathrm{p}}}\biggr)^{n_{\mathrm{s}}-1} \, (k_{0} \sigma_{\mathrm{rec}})^2 {\mathcal I}_{\ell}(n_{\mathrm{s}}),
\label{SP8}\\
 {\mathcal I}_{\ell}(n_{\mathrm{s}}) &=& \int_{0}^{\infty} x_{0}^{n_{\mathrm{s}} -4} j_{\ell}^2(x_{0}) \sin^2{[\gamma(\tau_{\mathrm{rec}}) x_{0}]} e^{ - 2 x_0^2/x_{\mathrm{D}}^2}\,\, d\,x_{0}.
 \label{SP8a}
\end{eqnarray}
where $\overline{C}^{(\mathrm{EE})}  = 3.5 \times 10^{-3} \, {\mathcal A}_{{\mathcal R}}$ and where  
\begin{equation}
\gamma(\tau_{\mathrm{rec}}) = \frac{1}{\tau_{0}} \int_{0}^{\tau_{\mathrm{rec}}} c_{\mathrm{sb}}(\tau) d\tau= \frac{\tau_{1}}{\sqrt{3}\tau_{0}}\int_{0}^{\tau_{\mathrm{rec}}/\tau_{1}} \frac{d x }{\sqrt{1 + \beta ( x^2 + 2 x)}}.
\label{SP9}
\end{equation}
The integral appearing in Eq. (\ref{SP8a}) can also be written as:
\begin{eqnarray}
&&{\mathcal I}_{\ell}(n_{\mathrm{s}},\ell_{\mathrm{D}}) = {\mathcal I}^{(1)}_{\ell}(n_{\mathrm{s}},\ell_{\mathrm{D}}) - 
{\mathcal I}^{(2)}_{\ell}(n_{\mathrm{s}},\ell_{\mathrm{D}}),
\label{SP11}\\
&& {\mathcal I}^{(1)}_{\ell}(n_{\mathrm{s}},\ell_{\mathrm{D}}) = \frac{1}{2}\ell^{n_{\mathrm{s}} -5} \int_{0}^{\infty} 
 (y^2 +1)^{(n_{\mathrm{s}} -6)/2} e^{- 2 (\ell^2/\ell_{\mathrm{D}}^2) (y^2 +1)}\,dy,
 \label{SP12}\\
 &&{\mathcal I}^{(2)}_{\ell}(n_{\mathrm{s}},\ell_{\mathrm{D}}) = \frac{1}{2}\ell^{n_{\mathrm{s}} -5} \int_{0}^{\infty} 
 (y^2 +1)^{(n_{\mathrm{s}} -6)/2} \cos{[2 \ell \gamma(\tau_{\mathrm{rec}})\sqrt{y^2 + 1}]}e^{- 2 (\ell^2/\ell_{\mathrm{D}}^2) (y^2 +1)}\, dy,
 \label{SP13}
 \end{eqnarray}
 It turns out that a good approximation to the full result can be obtained by setting 
 \begin{equation}
 C_{\ell}^{(\mathrm{EE})} = C_{\ell_{\mathrm{p}}}^{(\mathrm{EE})} \biggl(\frac{\ell}{\ell_{\mathrm{p}}}\biggr)^{n_{\mathrm{s}}-1 } e^{- 2 (\ell/\ell_{\mathrm{D}})^2},
 \label{SP14}
 \end{equation}
 where $\ell_{\mathrm{p}}$ is a given putative frequency. 
The estimate of the angular power spectrum of the E-mode can be improved by solving 
more accurately Eq. (\ref{DIP}) which can be written as 
\begin{equation}
\Delta_{\mathrm{I}1}(k,\tau) = -\,c_{\mathrm{sb}}^{3/2}[ a_{{\mathcal R}}(k,\tau) 
 - a_{\mathrm{B}}(k) ] \sin{[k r_{\mathrm{s}}(\tau)]}.
\label{SP15}
\end{equation}
where 
\begin{eqnarray}
&& a_{{\mathcal R}}(k,\tau) =  \frac{6}{25} {\mathcal R}_{*}(k)\ln{\biggl[\frac{14}{25} k\tau_{\mathrm{eq}}\biggr]} e^{- k^2/k_{\mathrm{D}}^2},
\label{SP16}\\
&& a_{\mathrm{B}}(k,\tau) = 3^{1/4} \biggl\{\biggl[\frac{\Omega_{\mathrm{B}}}{4}(k) - \sigma_{\mathrm{B}}(k)\biggr] + \frac{R_{\gamma}}{20} \Omega_{\mathrm{B}}(k)\biggr\} e^{- k^2/k_{\mathrm{D}}^2}.
\label{SP17}
\end{eqnarray}
These expressions allow for more accurate results. However because of the logarithmic factors
they are more difficult to handle inside the integrals. 
\subsection{B-mode autocorrelations}

Having found the E-mode polarization to zeroth order in the Faraday rotation rate,  the calculation can now 
be extended to first-order, i.e. when $\Delta_{+} \neq \Delta_{-}$.  As previously 
noticed \cite{gk3}, for magnetic fields of comoving intensity smaller than $10$ nG the Faraday autocorrelation can  be safely evaluated in the limit of small rotation 
angle\footnote{The rationale of this statement is very simple. If we consider the smallest 
imaginable frequency we can derive which is the bound on the comoving 
magnetic field over the pivot scale $k_{\mathrm{L}}^{-1}$ corresponding to a Faraday 
$|\Phi(\hat{n})|<1$. The result of this analysis \cite{gk3} imply, roughly, that for comoving 
frequencies of $30$ GHz, comoving fields of $100$ to $1000$ nG still lead to a rotation rate 
smaller than $1$. For larger frequencies the bound becomes looser. For a range 
of variation of $n_{\mathrm{B}}$ between, say, $1$ and $2.5$ the constraint changes by less than one order of magnitude.}. This will also be the strategy followed here so that, in real space, 
 the first-order iteration leads to an expected mismatch between the two aforementioned 
 orthogonal combinations:
\begin{equation}
\Delta_{+}(\hat{n},\tau) = - 2 \,\Phi(\hat{n},\tau)\, \Delta_{\mathrm{P}}(\vec{x},\hat{n},\tau), \qquad 
\Delta_{-}( \hat{n},\tau) =  2\, \Phi(\hat{n},\tau) \,\Delta_{\mathrm{P}}(\hat{n},\tau),
\label{BE10}
\end{equation}
where $\Delta_{\mathrm{P}}(\hat{n},\tau)$  follows from the solution of the heat transfer 
equation to zeroth order in the Faraday rate. 
From Eqs. (\ref{BE3}) and (\ref{BE4}) the B-mode polarization turns out to be given by:
\begin{eqnarray}
\Delta_{\mathrm{B}}(\vec{x},\hat{n},\tau) &=& 2 \, \partial_{\mu}^2[ (1 - \mu^2)\,\Phi(\vec{x},\hat{n},\tau)\,
 \Delta_{\mathrm{P}}(\hat{n},\tau)]
\label{BE11}\\
a_{\ell\, m}^{(\mathrm{B})} &=&\frac{1}{N_{\ell}}  \int d \hat{n}\, Y_{\ell m}^{*}(\hat{n}) \,\Delta_{\mathrm{B}}(\hat{n},\tau),
\label{BE12}
\end{eqnarray}
whose explicit expression can also be written as 
\begin{equation}
a_{\ell\, m}^{(\mathrm{B})}= \int\, d\hat{n} \,\, \Phi(\hat{n})\,\, \Delta_{\mathrm{P}}(\hat{n})\,\, [ _{2} Y_{\ell\, m}^{*}(\hat{n}) + _{-2} Y_{\ell\, m}^{*}(\hat{n})].
\label{BE13}
\end{equation}
For sufficiently small 
angular separations, the $2$-sphere degenerates into a plane where 
the flat basis is characterized 
by a pair of (Cartesian) coordinates $\vartheta_{x}$ and $\vartheta_{y}$. In the flat basis (as opposed to the spherical basis) 
the third direction is fixed.
The limit in which the 2-sphere degenerates into a plane is one of the simplest examples of group contraction \cite{vilenkin}. Owing to this observation the matrix elements of the representations 
of $SO(3)$ can be related to the matrix elements of the representations of $M(2)$ (i.e., in the 
language of \cite{vilenkin}, the $SO(2)$ group supplemented by the two-dimensional Euclidean translations). 
In the limit of large multipoles, therefore, the Legendre polynomials 
degenerate into Bessel functions, i.e., more precisely
\cite{vilenkin}:
\begin{equation}
\lim_{\ell \to \infty} \, P_{\ell}\biggl(\cos{\frac{r}{\ell}}\biggr) = J_{0}(r) \simeq J_{0}(\ell \vartheta) .
\label{BE14}
\end{equation}
At the right hand side $J_{0}(r)$ is a Bessel function of zeroth order while 
at the left-hand side it appears a Legendre polynomial whose argument can also be written as $P_{\ell}(\cos{\vartheta})$. The limit (\ref{BE14}) corresponds to keep fixed $\ell \vartheta$
by sending $\ell \to \infty$ and $\vartheta \to 0$. 

Every calculation conducted in the spherical basis can be performed in the flat basis and the results of the two calculations must coincide in the limit $\ell \to \infty$ and $\theta\to 0$. 
Consider, for instance, the temperature fluctuations computed 
in the spherical basis 
\begin{equation}
\Delta_{\mathrm{I}}(\vec{\vartheta}) = \frac{1}{2\pi} \int d^2\vec{\ell} \,
\Delta_{\mathrm{I}}(\vec{\ell})\, e^{i \vec{\ell} \cdot \vec{\vartheta}}, \qquad 
\langle \Delta_{\mathrm{I}}(\vec{\ell}) \Delta_{\mathrm{I}}(\vec{\ell}\,') \rangle = C^{(\mathrm{TT})}_{\ell}
\delta^{(2)}(\vec{\ell} + \vec{\ell}\,').
\label{BE15}
\end{equation}
A simple calculation in the flat basis shows that 
\begin{equation}
C^{\mathrm{TT}}(\vartheta_{12}) = \langle \Delta_{\mathrm{I}}(\vec{\vartheta_{1}}) \, \Delta_{\mathrm{I}}(\vec{\vartheta_{1}}) \rangle = \frac{1}{2\pi} \int\, \ell\,d\ell\,\, C_{\ell}^{\mathrm{TT}}\, J_{0}(\theta_{12} \, \ell) ,
\label{BE16}
\end{equation}
where the Bessel function arises from the integral 
\begin{equation}
\frac{1}{4\pi^2} \int d^2 \, \vec{\ell}\, C_{\ell}^{(\mathrm{TT})} \, e^{i \vec{\ell}\cdot 
(\vec{\vartheta}_{1} - \vec{\vartheta}_{2})}= 
\int \ell d\ell \int_{0}^{2\pi} \, d\varphi_{\ell} \,\, e^{i \ell \,\vartheta_{12}\, \cos{\varphi_{\ell}}} \equiv 
\frac{1}{2\pi} \int \ell d\,\ell\, C_{\ell}^{(\mathrm{TT})}\,\,J_{0}( \ell \vartheta_{12}).
\label{BE17}
\end{equation}
The second equality in Eq. (\ref{BE17}) follows from a well known integral representation 
of Bessel functions \cite{abr1,abr2}; the result of Eq. (\ref{BE17}) can also be
obtained from the temperature  autocorrelation computed in the spherical basis 
 by consistently taking the limit 
$\ell\to\infty$ and $\vartheta\to 0$:
\begin{equation}
C^{\mathrm{TT}}(\vartheta_{12}) = \sum_{\ell\,m} \frac{2\ell + 1}{4\pi}\, C_{\ell}^{(\mathrm{TT})} P_{\ell}(\cos{\vartheta_{12}}) \to  \frac{1}{2\pi} \int \ell\,d\ell\, C_{\ell}^{\mathrm{TT}} J_{0}(\theta_{12} \, \ell),
\label{BE18}
\end{equation}
where the second expression follows from Eq. (\ref{BE14}). 
Equations (\ref{BE15})--(\ref{BE18})  demonstrate 
the equivalence of the two computational schemes.
The polarization 
observables and the spin weighted spherical harmonics can be expressed in the flat basis \cite{sud,zalda} (see also 
\cite{moll,hu1} for a connection with the well known problem of computing ellipticity correlations from weak lensing in the flat basis).

For a semi-analytic estimate of the B-mode autocorrelation, Eq. (\ref{BE13}) can be inserted
inside Eq. (\ref{int5}). The derivation will be performed in the spherical basis, mentioning, when 
relevant, how the results also emerge by working in the flat basis.
Following the calculation originally due to Goldberg 
and Spergel \cite{GS} the B-mode autocorrelation can be recast in the form 
\begin{eqnarray}
C_{\ell}^{(\mathrm{BB})}&=& \sum_{j_{1}\, \, j_{2}} {\mathcal F}(\ell, j_{1}, j_{2}) \, C_{j_{2}}^{(\mathrm{EE})}(\ell_{\mathrm{D}},
\ell_{\mathrm{t}})\, 
C^{(\mathrm{FF})}_{j_{1}}(\ell_{\mathrm{D}},
\ell_{\mathrm{t}})
\nonumber\\
&\to& \int d j_{1}\, \int d j_{2}  {\mathcal F}(\ell, j_{1}, j_{2}) \, C_{j_{2}}^{(\mathrm{EE})}(\ell_{\mathrm{D}},
\ell_{\mathrm{t}})\, 
C^{(\mathrm{FF})}_{j_{1}}(\ell_{\mathrm{D}},
\ell_{\mathrm{t}}),
\label{AP1}
\end{eqnarray}
where sums have been replaced with integrals since, in the small-scale limit, $j_{1}$ and $j_{2}$ become, in practice, continuous variables. In Eq. (\ref{AP1}) the dependence 
of the angular power spectra  upon  $\ell_{\mathrm{t}}$ and 
$\ell_{\mathrm{D}}$ has been made explicit.
 The exact form of ${\mathcal F}(\ell,\,j_{1},\, j_{2})$ can be written as:
 \begin{equation}
 {\mathcal F}(\ell, j_{1}, j_{2})  = \frac{2j_{2}+1}{4\pi} 
 \biggl(\frac{2j_{1} + 1}{2\ell + 1}\biggr)\frac{f(\ell, j_{1}, j_{2})}{\ell \,(\ell -1)\, (\ell +1)\, (\ell +2) \,j_{2}\,(j_{2} -1)\, (j_{2} +1)\, (j_{2} +2)} \,\, |{\mathcal C}^{\ell \,\,0}_{j_1\,\,0\,\,j_{2}\,\,0}|^2,
 \label{AP2}
 \end{equation}
where 
\begin{eqnarray}
 f(\ell, j_{1}, j_{2}) &=& 
 2 j_{1} j_2(j_{1} +1) (j_{2} +1) + 2 j_1 \ell (j_{1} + 1) (\ell + 1)  - \ell^2 (\ell+1)^2 
 \nonumber\\
 &-& j_{1}^2 (j_{1}+1)^2 - j_{2}^2 (j_{2}+1)^2  + 2 j_{2}( j_{2}+1) + 2 \ell (\ell +1)- 2 j_{1}(j_{1} +1),
\label{AP3}\\
 |{\mathcal C}^{\ell \,\,0}_{j_1\,\,0\,\,j_{2}\,\,0}|^2 &=& \frac{2 \ell +1}{4\pi} \int_{-1}^{1} \, d\mu\,P_{\ell}(\mu)\, P_{j_{1}}(\mu) \,P_{j_{2}}(\mu).
\label{AP4}
\end{eqnarray}
In quantum mechanics, 
the Clebsch-Gordon coefficients are  probability amplitudes arising when we have two different 
angular momenta (be them $\vec{J}_{1}$ and $\vec{J}_{2}$) acting on two different Hilbert spaces. 
The complete and orthonormal set of eigenvectors of $J_{1}^2 $ and of $J_{1\,z}$  are  $|j_{1} \, m_{1}\rangle$.
Similarly, the  complete and orthonormal set of eigenvectors 
of $J_{2}^2 $ and of $J_{2\,z}$  are  $|j_{2} \, m_{2}\rangle$. By doing the tensor product between the vectors 
of the two subspaces with $(2 j_{1} + 1)$ and $(2 j_{2} +1)$ dimensions we obtain the decoupled representation
whose basis vectors are characterized by four different eigenvalues, i.e. $j_{1}$, $j_{2}$, $m_{1}$ and $m_{2}$.
The same problem can be addressed by exploiting the coupled representation where the eigenvectors are characterized 
by a different system of eigenvalues i.e. $\ell$, $m$ together with $j_{1}$ and $j_{2}$. 
Since both systems of eigenvectors are complete and orthonormal they are connected by a unitary 
transformation 
\begin{eqnarray}
&& |j_{1}\,j_{2}\,\, \ell\, m\rangle = \sum_{m_1,\,\,m_2}  \,\, {\mathcal C}^{\ell \,\,m}_{j_1\,m_{1}\,\,j_{2}\,m_{2}} 
  |j_{1}\,j_{2}\,\, m_{1}\, m_{2} \rangle,
\nonumber\\  
 && |{\mathcal C}^{\ell \,\,m}_{j_1\,m_{1}\,\,j_{2}\,\,m_{2}}|^2  = |\langle j_{1}\, 
  j_{2}\,\,m_{1}\, m_{2}| j_{1}\,j_{2}\,\,\ell\, m\rangle|^2, \qquad |j_{1} - j_{2}| \leq\ell \leq j_{1} + j_{2},
\label{AP5}
 \end{eqnarray} 
 reminding that the  Clebsch-Gordon coefficients vanish unless the triangle inequality is obeyed. 
From  Eq. (\ref{AP5}) it can be easily argued that
${\mathcal C}^{\ell \,\,m}_{j_1\,m_{1}\,\,j_{2}\,\,m_{2}}$ 
is the $3j$ Wigner symbol up to a numerical factor 
(i.e. $\sqrt{2\ell +1}$) and up to a phase (i.e. $(-1)^{j_{1}-j_{2} + m_{3}}$)\cite{edmonds}. Furthermore, from Eq. (\ref{AP5}),
${\mathcal C}^{\ell \,\,0}_{j_1\,0\,\,j_{2}\,0}$ is zero if the sum of the eigenvalues gives an odd integer and 
and it is different from zero otherwise. All the mentioned properties can also deduced 
in the particular case of Eq. (\ref{AP4}) and bearing in mind the recurrence relations of the Legendre 
polynomials.
The limit where $j_{1}\gg 1$, $j_{2} \gg 1$ and $\ell \gg 1$ is physically opposite to the ``quantum" limit where 
Eq. (\ref{AP5}) originally arises. We could say that, using the quantum mechanical analogy, the small-scale 
limit is ``classical" since there the Clebsch-Gordon coefficients can be thought to arise 
from the composition of classical angular momenta. This approach to the asymptotics of the Clebsch-Gordon 
coefficients was originally studied by Ponzano and Regge \cite{ponzano} by exploiting 
the connection of the Clebsch-Gordon coefficients with the Wigner $3j$ and $6j$ symbols. In the case which is 
mostly relevant for the present investigation we have that \cite{ponzano} (see also \cite{gordon1,gordon2})
\begin{equation}
|{\mathcal C}^{\ell \,\,0}_{j_1\,\,0\,\,j_{2}\,\,0}|^2 = \frac{2 \ell + 1}{\pi} \frac{1}{\sqrt{4 j_{1}^2 j_{2}^2 - (\ell^2 - j_{1}^2 - j_{2}^2)^2}}.
\label{AP6}
\end{equation}
It is worth mentioning that, in the general case, the semiclassical limit of the Clebsch-Gordon coefficients 
\cite{gordon1} (see also \cite{gordon2}) has been studied with a technique which  is closely related to the 
perspective adopted in the present investigation. Indeed in \cite{gordon1}  the results of \cite{ponzano} have been confirmed \footnote{It must be said that the discussion 
of \cite{ponzano} was, in its nature, rather formal. The authors defined asymptotic expressions valid in 
quantum, classical and semi-classical domains without 
discussing the smooth connections of the different 
limits. This aspect has been later deepened in \cite{gordon1,gordon2}. This is why it will be important 
to cross-check the obtained results with direct numerical evaluations of the relevant matrix elements. This will be 
one of the themes of the following section.}
by using the recursion relations obeyed by the $3j$ and $6j$ symbols. The latter relations become, in the 
semiclassical limit, differential equations. In the present approach we somehow use the same observation 
when replacing discrete sums by integrals (see Eq. (\ref{AP1})). 
In the flat basis, the result of Eq. (\ref{AP6}) follows from  
the evaluation of the integral 
\begin{equation}
\ell \int_{0}^{\infty} 
J_{0}(\ell_{1}\,y) J_{0}(\ell_{2}\,y) J_{0}(\ell\, y) y d y.
\label{AP6b}
\end{equation}
which coincides exactly with the right-hand side of Eq. (\ref{AP6}) (see, e.g. \cite{vilenkin,grad})
when $|\ell_{1} - \ell_{2}| < \ell < \ell_{1} + \ell_{2}$ and it is zero otherwise.
  Inserting Eqs. (\ref{AP3}), (\ref{AP4}) and (\ref{AP6}) into Eq. (\ref{AP2})  we do get, after simple but lengthy algebra, that 
\begin{eqnarray}
{\mathcal F}(\ell,\, x,\, y) &=& \frac{\left( x + y \right) \,{\left[  \ell^2\,{\left( x + y \right) }^2  - 2\,x^2\,y^2  -2\,\ell^4\right] }^2}
  {2 \pi^2\,\,\ell^4\,\,{\left( x - y \right) }^3\,{\sqrt{\left(  x -\ell \right) \,\left(  x + \ell \right) \,\left( \ell - y \right) \,\left(\ell + y \right) }}} 
  \label{AP7a}\\
  &\to& \frac{\left( \tilde{x} + \tilde{y} \right) \,{\left[  \,{\left( \tilde{x} + \tilde{y} \right) }^2  - 2\,\tilde{x}^2\,\tilde{y}^2  -2\,\right] }^2}
  {2 \pi^2\,\,\,\,{\left( \tilde{x} - \tilde{y} \right) }^3\,{\sqrt{\left(  \tilde{x} -1 \right) \,\left(  \tilde{x} + 1 \right) \,\left( 1 - \tilde{y} \right) \,\left( 1+ \tilde{y} \right) }}}.
\label{AP7}
\end{eqnarray}
In Eq. (\ref{AP7a}), for immediate convenience, the following change of variables has been adopted
\begin{equation}
x = j_{1} + j_{2},\qquad y = j_{1} - j_{2},\qquad 0< y < \ell , \qquad x> \ell,
\label{AP8}
\end{equation}
where the range of variation of $x$ and $y$ follows from the triangle inequality. In Eq. (\ref{AP7}) a further change of variable has been performed, i.e. $x\to \tilde{x} = x/\ell$ and 
$y\to \tilde{y} = y/\ell$. In terms of $\tilde{x}$ and $\tilde{y}$ the integration region becomes 
$0 < \tilde{y} < 1$ and $\tilde{x} > 1$. 

The largest signal obtainable from Faraday rotation arises for frequencies smaller than the maximum of the CMB (i.e., roughly, $\nu < \nu_{\mathrm{max}} \simeq 222 \,\, \mathrm{GHz}$) 
and for angular scales much smaller than one degree. Over these scales 
$C_{\ell}^{(\mathrm{BB})}$ increases with $\ell$. Recent experiments Quad \cite{quad1,quad2,quad3} give 
upper limits on the B-mode polarization also for $\ell \gg 200$ and, with large error bars, up to $\ell \simeq 1987$.
The autocorrelations of the polarization and of the Faraday 
rotation can be expressed, respectively, as 
\begin{eqnarray}
C_{\ell}^{(\mathrm{FF})} = {\mathcal N}_{\mathrm{F}} \,\, \ell^{n_{\mathrm{B}} -3} {\mathcal G}(\ell/\ell_{\mathrm{d}}, n_{\mathrm{B}})  e^{- 2\ell^2/\ell_{\mathrm{d}}^2},
\qquad C_{\ell}^{(\mathrm{EE})} = {\mathcal N}_{\mathrm{E}} \,\, \ell^{n_{\mathrm{s}} -1} {\mathcal G}(\ell/\ell_{\mathrm{D}}, n_{\mathrm{s}})  e^{- 2\ell^2/\ell_{\mathrm{D}}^2},
\label{AP9}
\end{eqnarray}
in the small scale limit.
The results of Eqs. (\ref{AP8}) and (\ref{AP9}) then allow for the following estimate 
of the B-mode autocorrelation
\begin{equation}
\frac{\ell (\ell + 1)}{2\pi} C_{\ell}^{(\mathrm{BB})} \simeq {\mathcal N}_{\mathrm{B}}
\biggl(\frac{\ell}{\overline{\ell}_{\mathrm{d}}}\biggr)^{(n_{\mathrm{B}} + n_{\mathrm{s}} + 2)/2}  
\, e^{ - \alpha_{1} (\ell/\overline{\ell}_{\mathrm{d}})^2}
\label{AP10}
\end{equation}
where $\alpha_{1} = 1.13$ and where 
\begin{eqnarray}
\overline{\ell}_{\mathrm{d}}^{-2} &=&  \ell_{\mathrm{D}}^{-2} + \ell_{\mathrm{d}}^{-2} \equiv  2 \ell_{\mathrm{D}}^{-2} + \ell_{\mathrm{t}}^{-2} + \ell_{\sigma}^{-2},
 \label{AP11}\\
 {\mathcal N}_{\mathrm{B}} &=& \frac{\sqrt{2}}{\pi^2} {\mathcal N}_{\mathrm{F}} \, {\mathcal N}_{\mathrm{E}} \, \overline{\ell}_{\mathrm{d}}^{n_{\mathrm{B}} + n_{\mathrm{s}}} \,\,\frac{\Gamma\biggl(\frac{5 - n_{\mathrm{B}}}{2} \biggr)
\,\Gamma\biggl(\frac{5 - n_{\mathrm{s}}}{2} \biggr)}{\Gamma\biggl(\frac{6 - n_{\mathrm{B}}}{2}\biggr) \, \Gamma\biggl(\frac{6 - n_{\mathrm{s}}}{2}\biggr)},
\label{AP12}\\
{\mathcal N}_{\mathrm{F}} &=& 5.075\times 10^{-6}\, \, \biggl(\frac{\sqrt{A}_{\mathrm{B}}}{\mathrm{nG}} \biggr)^2 \biggl(\frac{\nu}{\nu_{\mathrm{max}}}\biggr)^{-4} \, \biggl(\frac{k_{0}}{k_{\mathrm{L}}}\biggr)^{n_{\mathrm{B}} -1},
\label{AP13}\\
{\mathcal N}_{\mathrm{E}} &=& 2.96 \times 10^{-4} \,\,\biggl(\frac{{\mathcal A}_{\mathcal R}}{2.41 \times 10^{-9}} \biggr) \,
\biggl(\frac{k_{0}}{k_{\mathrm{p}}}\biggr)^{n_{\mathrm{s}}-1}\,\,
\biggl(\frac{T_{\gamma0}}{2.725}\biggr)^2\,\, (\mu\,\mathrm{K})^2.
\label{AP14}
\end{eqnarray}
 The result of  Eq. (\ref{AP10}) follows from the evaluation of the integral:
\begin{eqnarray}
&&{\mathcal W}(u, v,n_{\mathrm{B}}, n_{\mathrm{s}}) = \int_{1}^{\infty} d \tilde{x} \int_{0}^{1} d\tilde{y} (\tilde{x} + \tilde{y})^{n_{\mathrm{B}}-2}\,(\tilde{x} - \tilde{y})^{n_{\mathrm{s}}-4}\,\times
\nonumber\\
&&\times \frac{
[(\tilde{x} + \tilde{y})^2 - 2 \tilde{x}^2 \tilde{y}^2 - 2]^2}{\sqrt{(\tilde{x} -1) (\tilde{x} +1) ( 1 - \tilde{y}) ( 1 + \tilde{y})}} e^{ - u^2( \tilde{x} + \tilde{y})^2/2}\,\, e^{ - v^2 (\tilde{x} - \tilde{y})^2/2},
\label{AP15}
\end{eqnarray}
 where $u = \ell/\ell_{\mathrm{d}}$ and $v = \ell/\ell_{\mathrm{D}}$. As indicated ${\mathcal W}(u, v,n_{\mathrm{B}}, n_{\mathrm{s}})$
 depends upon four parameters, i.e. $n_{\mathrm{B}}$, $n_{\mathrm{s}}$, $u$ and $v$. Since we will be mainly interested 
 in the case where $n_{\mathrm{s}}$ is closed to its best-fit value we shall assume that $n_{\mathrm{s}}$ varies 
 between $0.9$ and $1.1$. The allowed variation on $n_{\mathrm{B}}$ must necessarily be 
 broader and it will be assumed, quite generally, that $0< n_{\mathrm{B}} < 5$ which include 
 the case of red (i.e. $n_{\mathrm{B}} <1$), scale-invariant (i.e. $n_{\mathrm{B}} \simeq 1$) and 
 blue (i.e. $n_{\mathrm{B}} > 1$) magnetic spectra. 
 Finally one can also assume that $u$ and $v$ are of the same oder, i.e. they can be different 
 but their mutual deviations are well within the order of magnitude and, typically, even less than a 
 factor of $2$. In this restricted parameter space the full integral of Eq. (\ref{AP15}) can be reasonably 
 approximated by the following expression: 
 \begin{equation}
 {\mathcal W}(u, v,n_{\mathrm{B}}, n_{\mathrm{s}}) \simeq 2^{n_{\mathrm{B}} + n_{\mathrm{s}} + 3/2} 
 ( u^2 + v^2)^{1/2 - n_{\mathrm{s}}/4 - n_{\mathrm{B}}/4}\,\, e^{ - \alpha_{1} (u^2 + v^2)}, \qquad 10^{-3} <\,u, v\,< 10.
 \label{AP16}
\end{equation}
 Note that in Eq. (\ref{AP16}) the range of variation of $u\simeq v$ is rather broad and much larger than what is needed. 
 Equation (\ref{AP16}) is certainly rather handy and shall be used, in a moment in direct estimate. 
 However, an even better analytical approximation can be obtained by integrating directly Eq. (\ref{AP15}) 
 for a given set of parameters. 
Before switching to the phenomenological implications of the present results, it is 
relevant to remind that the present estimates are all conducted in the framework 
of the (minimal) $\Lambda$CDM scenario, in this sense the restrictions on the variation 
of the scalar spectral index are fully justified.  
\renewcommand{\theequation}{5.\arabic{equation}}
\setcounter{equation}{0}
\section{Parameter space and parameter extraction}
\label{sec5}
The expression of Eq. (\ref{AP10}) attains its maximum for 
\begin{equation}
\ell_{\mathrm{max}} =\frac{1}{2}\sqrt{\frac{n_{\mathrm{s}} + n_{\mathrm{B}} + 2}{\alpha_{1}}} \,\,\overline{\ell}_{\mathrm{d}},
\label{PH1}
\end{equation}
In Eqs. (\ref{AP12}) and (\ref{AP13}), following the WMAP5-yr data alone, $k_{0} = 7.08 \times 10^{-5}\,\, \mathrm{Mpc}^{-1}$. The value of the pivot scale $k_{\mathrm{p}}$ is set, as usual 
to $0.002\,\,\mathrm{Mpc}^{-1}$. The value of the magnetic pivot scale $k_{\mathrm{L}}$ will be taken to be $1\,\mathrm{Mpc}^{-1}$. 
The estimates of the previous sections will give, for $\overline{\ell}_{\mathrm{d}}$, a value laying between 
$\ell \simeq 900$ and $\ell \simeq 1100$. In the case of the best fit parameters 
of the WMAP5 data alone we will have that $\overline{\ell}_{\mathrm{d}} \simeq 911$.
\begin{figure}[!ht]
\centering
\includegraphics[height=6.3cm]{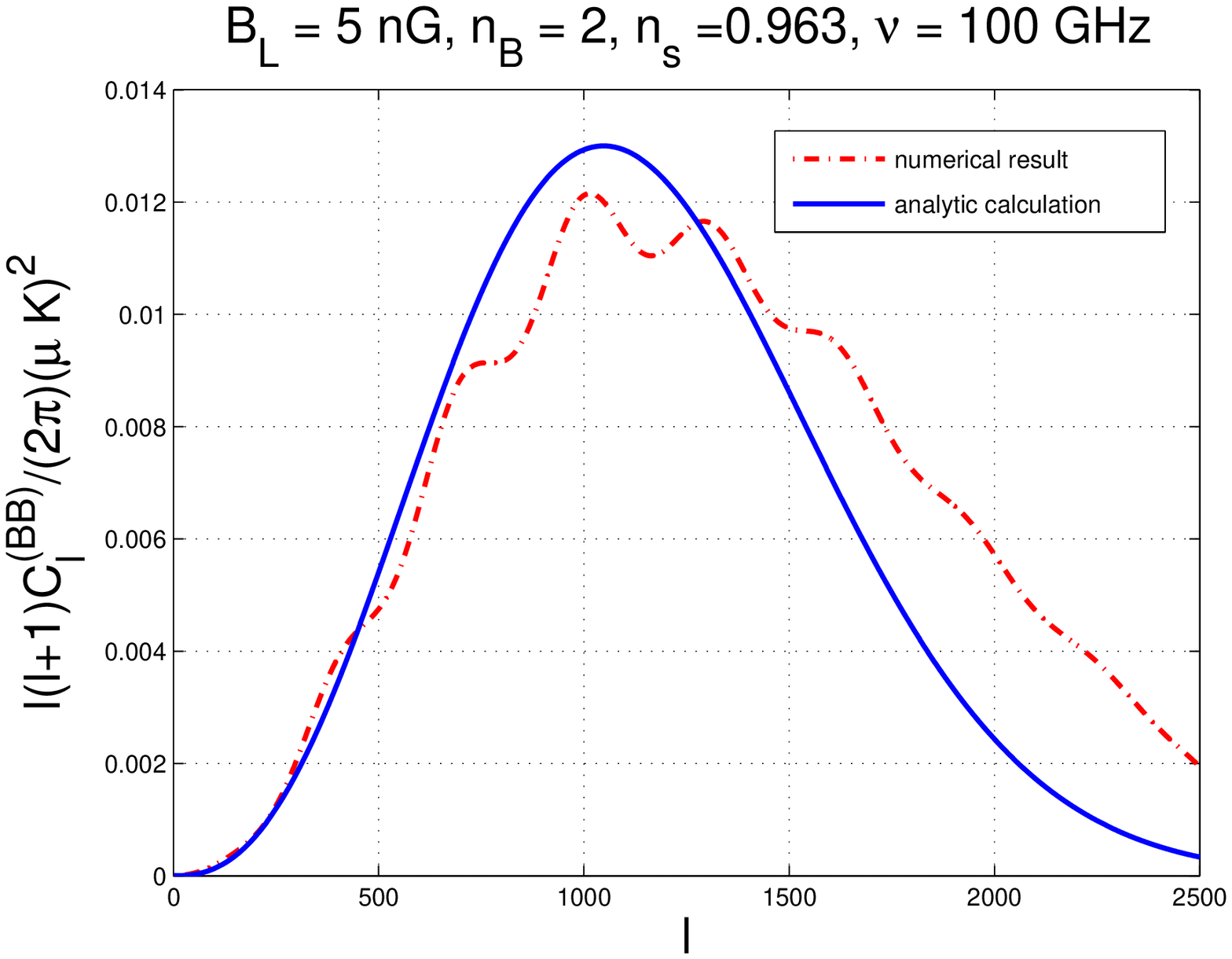}
\includegraphics[height=6.3cm]{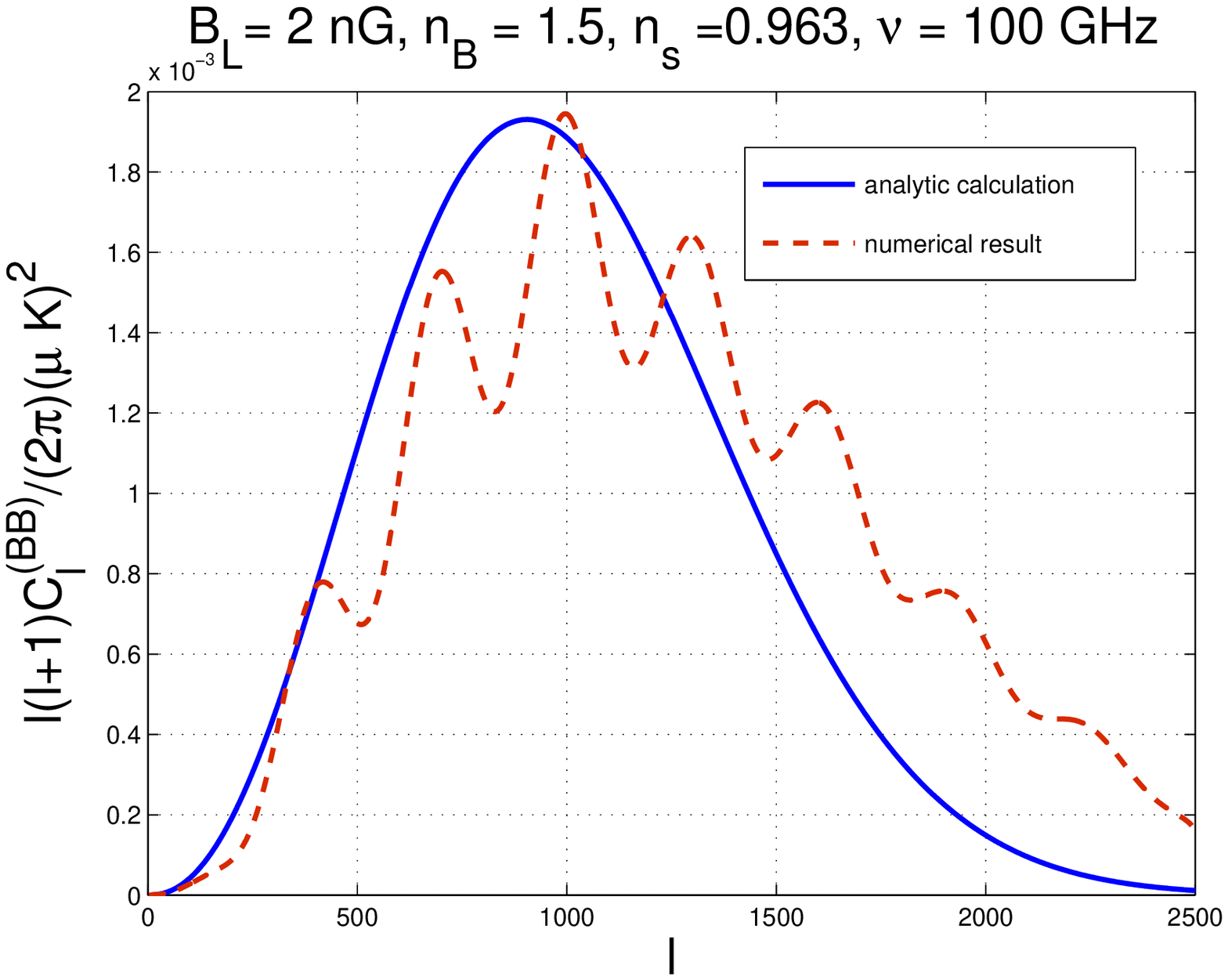}
\caption[a]{Comparison between the analytic and numerical estimates of the magnetized B-mode 
autocorrelation.}
\label{Figure2}      
\end{figure}
In Fig. \ref{Figure2} the full numerical calculation is compared with the analytic results obtained in the previous 
section. In the plot at the left the magnetic field has been set to $B_{\mathrm{L}} = 5$ nG (with $n_{\mathrm{B}} = 2$) while 
in the plot at the right $B_{\mathrm{L}} = 2$ nG (with $n_{\mathrm{B}} = 1.5$); $B_{\mathrm{L}}$ is the 
value of the magnetic energy density regularized over a spatial domain $L\simeq k_{\mathrm{L}}^{-1}$ where 
$k_{\mathrm{L}} = \mathrm{Mpc}^{-1}$ is the magnetic pivot scale introduced in Eq. (\ref{EQ2}). The regularization scheme depends upon the spectral slope: for $n_{\mathrm{B}} >1$ (blue spectra) it 
is often performed by means of a Gaussian window function. In the case of red spectra 
(i.e. $n_{\mathrm{B}}<1$) the regularization is done  by means of a smooth representation of a step function. 
The relation between $B_{\mathrm{L}}^2$ and $A_{\mathrm{B}}$ is given by 
\begin{eqnarray}
A_{\mathrm{B}} &=& \frac{ (2\pi)^{n_{\mathrm{B}}-1}}{\Gamma\biggl(\frac{n_{\mathrm{B}} -1}{2}\biggr)}\,\, B_{\mathrm{L}}^2,\qquad 
n_{\mathrm{B}} >1,
\label{PH3}\\
A_{\mathrm{B}} &=& \frac{1 - n_{\mathrm{B}}}{2} \biggl(\frac{k_{0}}{k_{\mathrm{L}}}\biggr)^{1 - n_{\mathrm{B}}}\,\, B_{\mathrm{L}}^2,
\qquad n_{\mathrm{B}} < 1.
\label{PH4}
\end{eqnarray}
The passage from $A_{\mathrm{B}}$ (which has dimensions of $\mathrm{nG}^2$) to $B_{\mathrm{L}}^2$ can be done at any stage of a given calculation and it is 
not really crucial. For instance, 
in the case of curvature perturbations the fit to the experimental data is customarily performed 
in terms of ${\mathcal A}_{{\mathcal R}}$ which is the analog of $A_{\mathrm{B}}$ and not of 
$B_{\mathrm{L}}^2$.  
In both plots of Fig. \ref{Figure2} the full line is obtained from Eq. (\ref{AP9}). The dashed and dot-dashed 
lines are instead the result of a direct numerical integration using the code developed in \cite{gk1,gk2,gk3}. 
In the plot at the left the theoretical result has been divided by a factor 2 while in the plot at the right 
the result of Eq. (\ref{AP9}) has been multiplied by a factor 2. This is due to the fact 
that the integral of Eq. (\ref{AP14}) has been approximated with Eq. (\ref{AP15}). In other 
words the approximation of Eq. (\ref{AP15}) is rather good for the slope at the expenses 
of a factor of 2 in the amplitude. The latter statement can be better understood 
by noticing that the position of the maximum is always well determined by Eq. (\ref{PH1}) 
and this reflects the correctness of the $\ell$ dependence in Eq. (\ref{AP9}). 
Equations (\ref{AP14}) and (\ref{AP15}) are quite useful if we wish 
to set direct bounds on the magnetic field intensity starting from published upper limits on the B-mode 
polarization.  

We are now going to compare the obtained results
 with the direct bounds on the B-mode autocorrelations.  There are  two complementary ways of making use of the data 
illustrated in Tab. \ref{TABLE1}. The first strategy is direct, i.e. from the upper limits on the B-mode 
autocorrelations, a constraint on the properties of the magnetic field intensity can be inferred. 
The second strategy is to take the data of each experiment at face value and to perform 
a global $\chi^2$ analysis. 
\begin{figure}[!ht]
\centering
\includegraphics[height=6.3cm]{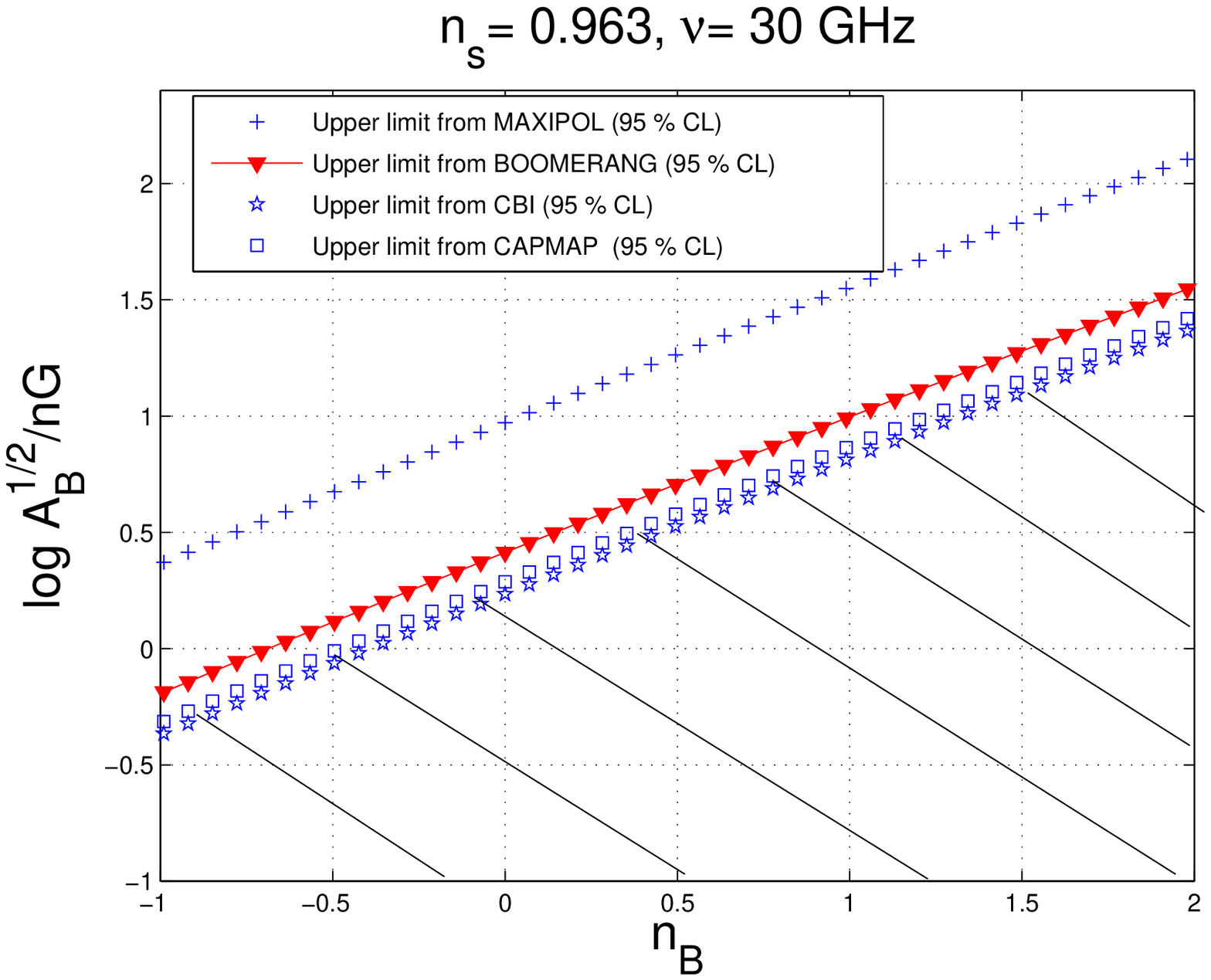}
\includegraphics[height=6.3cm]{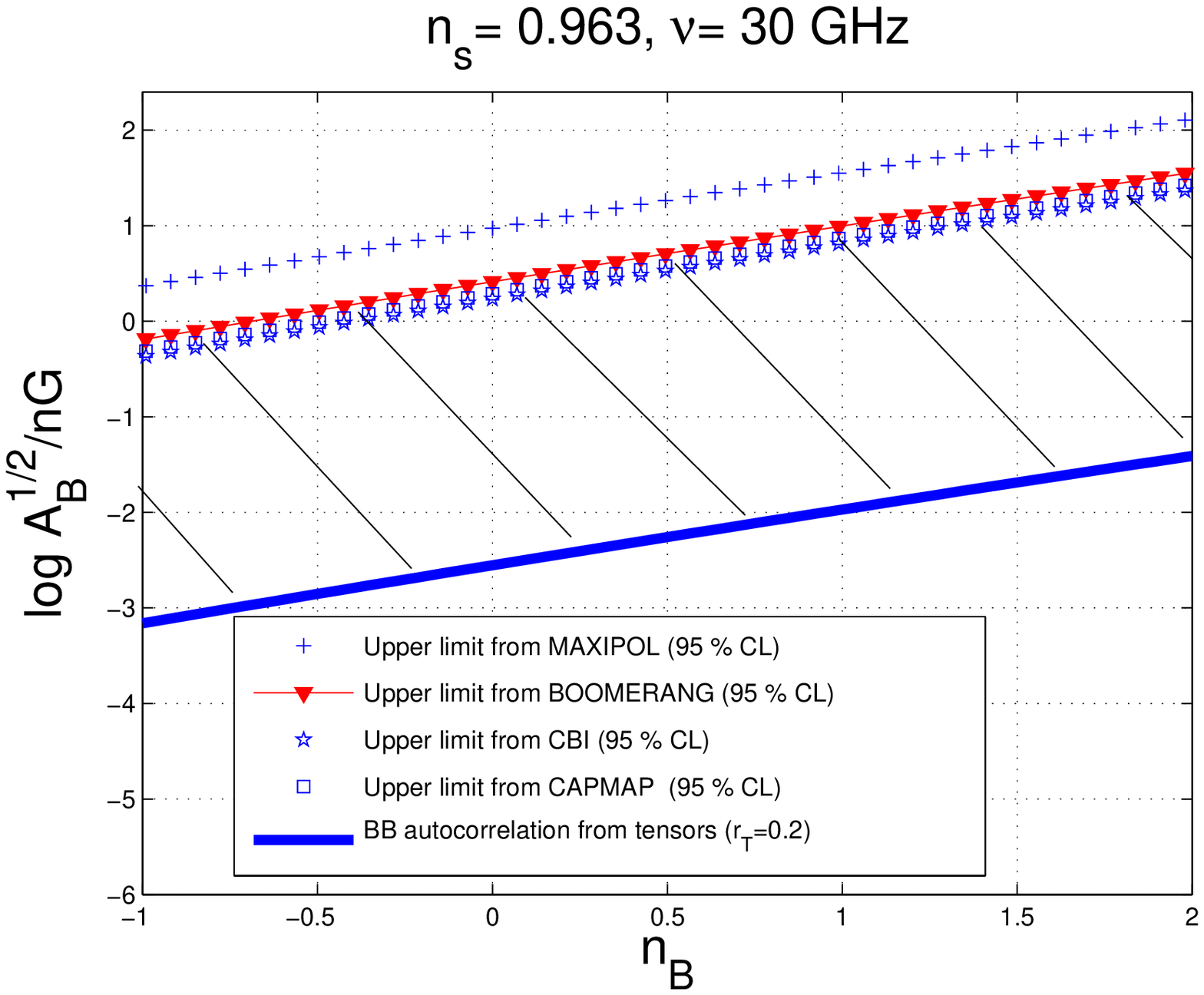}
\caption[a]{Bounds on the parameter space of the magnetized background from Maxipol, Boomerang, Cbi and Capmap. The full line in the right plot denotes the foreseen signal for a stochastic background 
of relic gravitons with $r_{\mathrm{T}} = 0.2$. The shaded area denote allowed regions 
of the parameter space.}
\label{Figure3}      
\end{figure}
 Four of the experiments of  Tab. \ref{TABLE1} reported direct upper limits 
on the B-mode autocorrelation. Maxipol\footnote{Maxipol is an acronym for 
``Millimiter Anisotropy experiment Imaging array''}\cite{maxip1,maxip2} and Boomerang\footnote{Boomerang is ana acronym for ``Balloon Observations of Millimetric Extragalactic Radiation and Geophysics"}
 \cite{boom1} derived, respectively, the 
following upper limits on the B-mode autocorrelation:
\begin{equation}
\frac{\ell(\ell+1)}{2\pi} C_{\ell}^{(\mathrm{BB})} \leq 112.36 \,\,(\mu\,\mathrm{K})^2,\qquad 
\frac{\ell(\ell+1)}{2\pi} C_{\ell}^{(\mathrm{BB})} \leq 8.6  \,(\mu\,\mathrm{K})^2,
\label{PH5}
\end{equation}
where both figures are at $95\%$ confidence level. The Maxipol limit becomes more tight (i.e. 
from $112.36\,\,(\mu\,\mathrm{K})^2$ down to  $90.25\,\,(\mu\,\mathrm{K})^2$) if the calibration uncertainty is excluded \cite{maxip1,maxip2}.  Both Boomerang and Maxipol 
operate at a relatively high frequency (i.e. $140$ GHz for Maxipol and $145$ GHz for Boomerang). 
Cbi\footnote{Cbi is an acronym for ``Cosmic Background Imager"} \cite{cbi} and Capmap\footnote{Capmap 
is the acronym for ``Cosmic Anisotropy Polarization Mapper".}\cite{capmap} gave more stringent 
upper limits, i.e., respectively
\begin{equation}
\frac{\ell(\ell+1)}{2\pi} C_{\ell}^{(\mathrm{BB})} \leq 3.76 \,\,(\mu\,\mathrm{K})^2,\qquad 
\frac{\ell(\ell+1)}{2\pi} C_{\ell}^{(\mathrm{BB})} \leq 4.8\,\,(\mu\,\mathrm{K})^2.
\label{PH6}
\end{equation}
The upper limits of Eq. (\ref{PH6}) are more stringent, in absolute value, than the ones of Eq. (\ref{PH5}). 
Furthermore, as discussed in the previous section, the magnetized B-mode autocorrelation 
scales with the frequency. Therefore the signal is larger for lower frequencies. The 
Cbi and the Capmap experiment are both working at frequencies lower than Maxipol and Boomerang. 
More precisely Cbi operates between $26$ and $36$ GHz. In the case of Capmap twelve receivers 
operate, approximately, in the W band (i.e. for $84\,\mathrm{GHz} <\nu < 100\,\mathrm{GHz}$) 
while the remaining four receivers operate in the so called Q band (i.e. for $35\,\mathrm{GHz} <\nu < 46\,\mathrm{GHz}$).  
\begin{figure}[!ht]
\centering
\includegraphics[height=6.3cm]{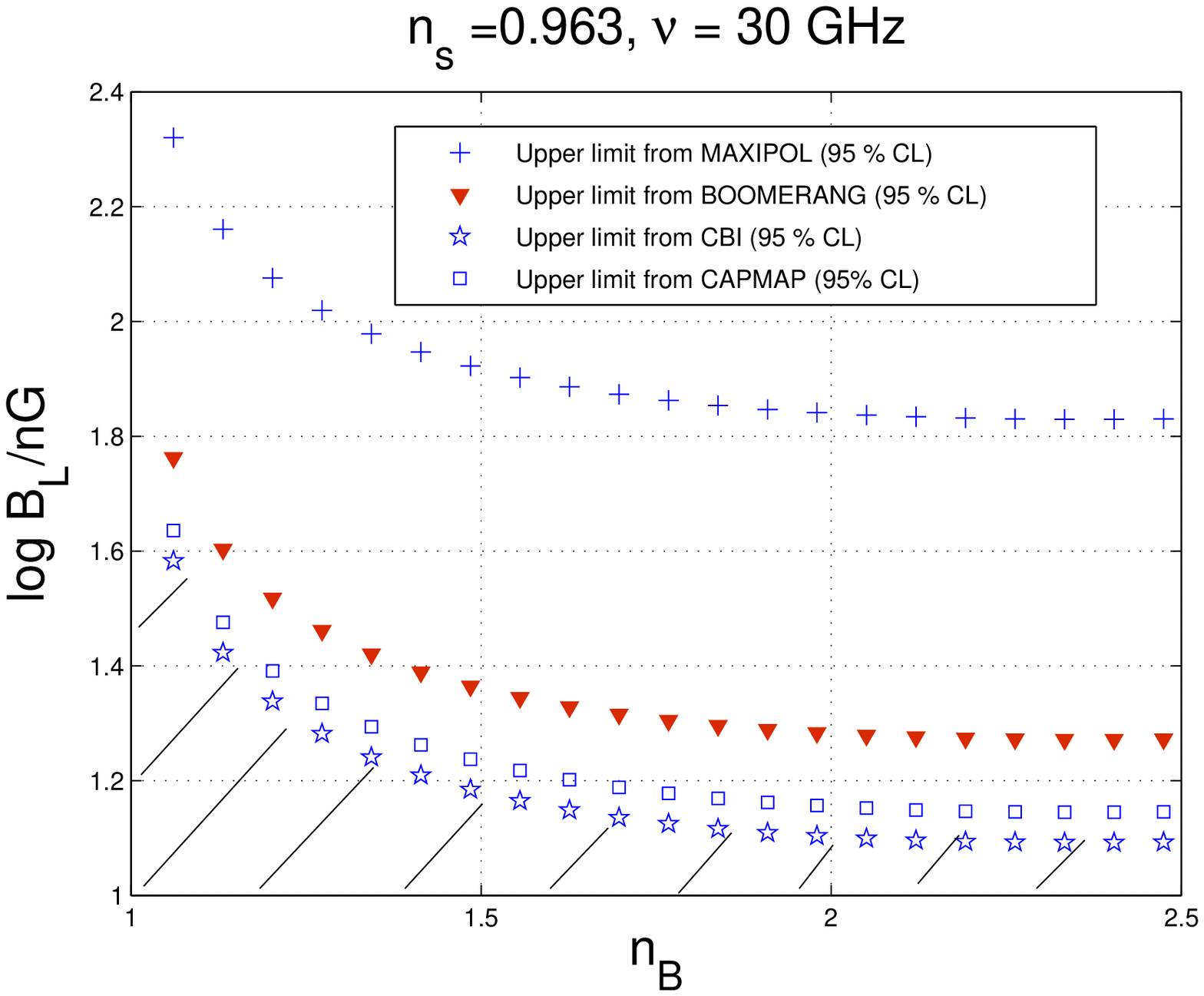}
\includegraphics[height=6.3cm]{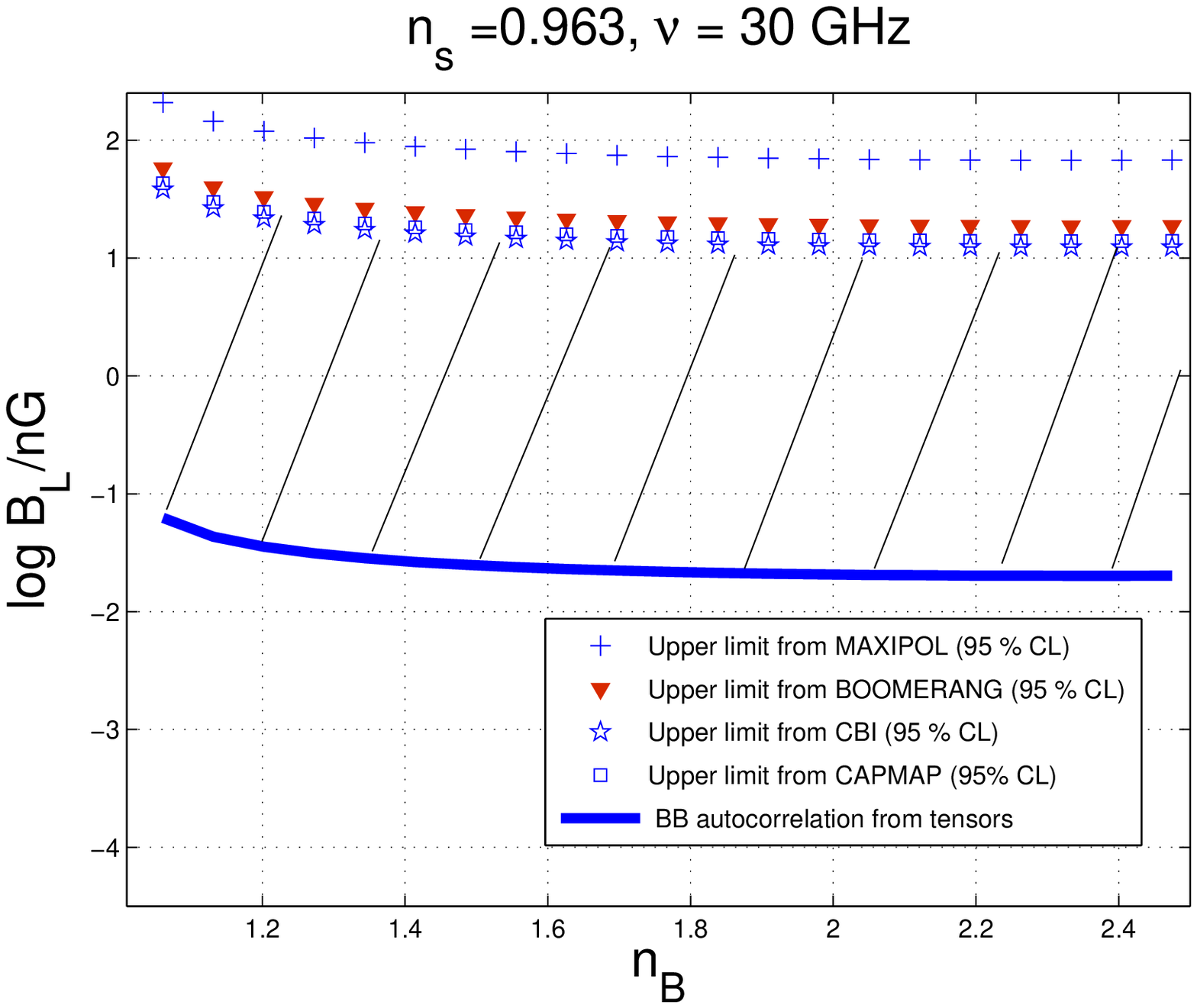}
\caption[a]{The same bounds illustrated in Fig. \ref{Figure3} are here illustrated in terms of 
$B_{\mathrm{L}}$ and for blue spectral index. As in Fig. \ref{Figure3} the shaded 
areas denote allowed regions of the parameter space. 
As in Fig. \ref{Figure3} the region below the full (thick) line in the plot at the right is not excluded but it would 
correspond to rather optimistic sensitivities 
of the forthcoming experiments.}
\label{Figure4}
\end{figure}
In Fig. \ref{Figure3} the bounds stemming from Eqs. (\ref{PH5}) and (\ref{PH6}) are 
compared with the signal of the magnetized B-mode. Concerning the exclusion plots 
of Fig. \ref{Figure3} few remarks are in order. 
Since the experiments gave specific upper limits, they must be applied, by definition, where 
the signal is larger and this occurs, in our case, for low frequencies and for $\ell \sim \ell_{\max}$.
The common putative frequency has been chosen, for the purposes of Fig. \ref{Figure3}, to be 
$30$ GHz. Both plots of Fig. \ref{Figure3} contain the same amount of informations. In the 
plot at the right the bottom line denotes the equivalent signal on the B-mode autocorrelation
in the case of a stochastic background of gravitons with tensor to scalar ratio $r_{\mathrm{T}} \simeq 0.2$.
The latter figure is typically one of the goals of forthcoming experimental programs aimed 
at observing the B-modes arising from the tensor modes  of the geometry. 
In this sense the shaded area allows to estimate the 
region of the parameter space which will be eventually accessible 
when the given experiment will improve their sensitivities 
by attaining progressively lower values in $r_{\mathrm{T}}$.      
\begin{figure}[!ht]
\centering
\includegraphics[height=6.3cm]{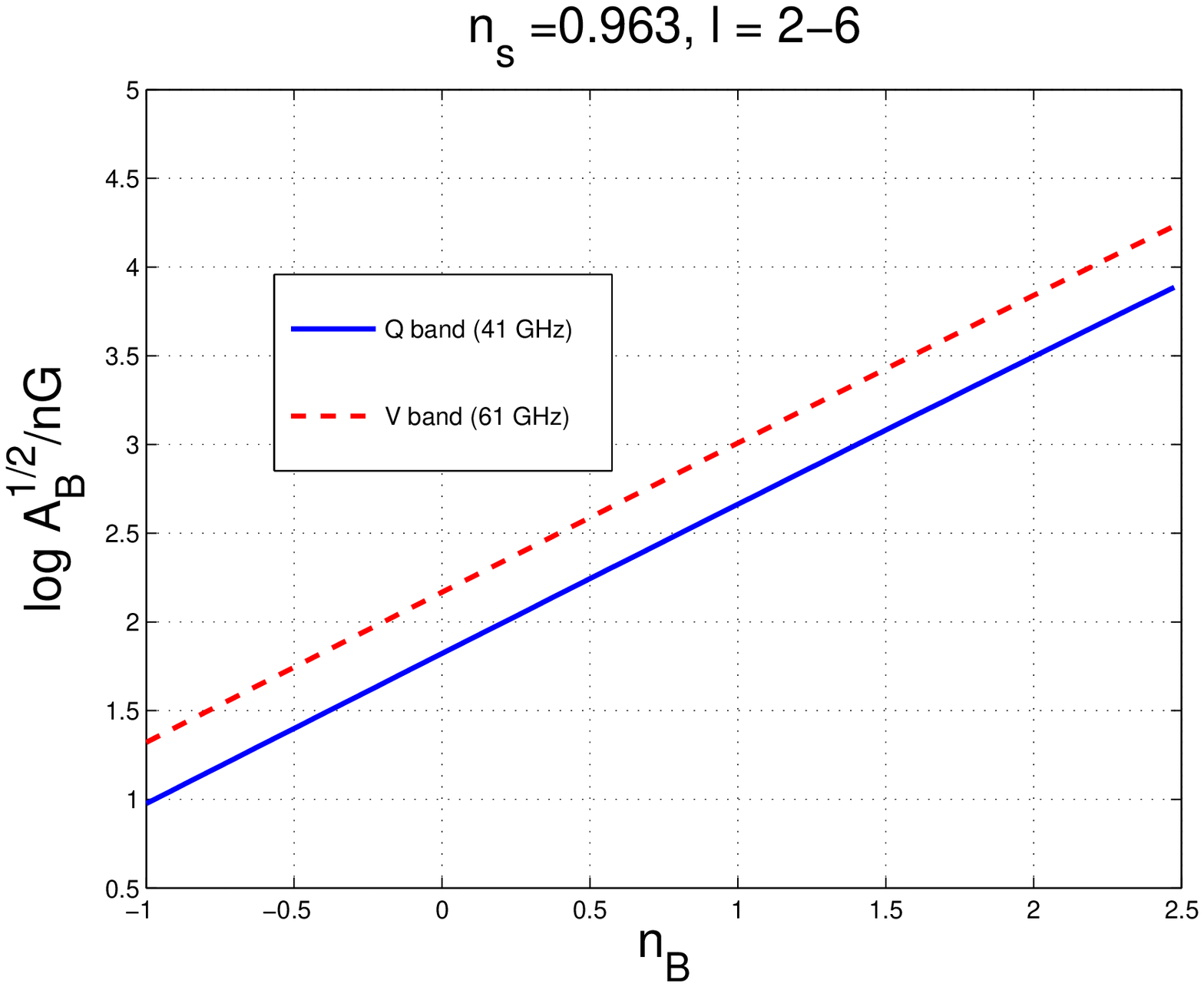}
\includegraphics[height=6.3cm]{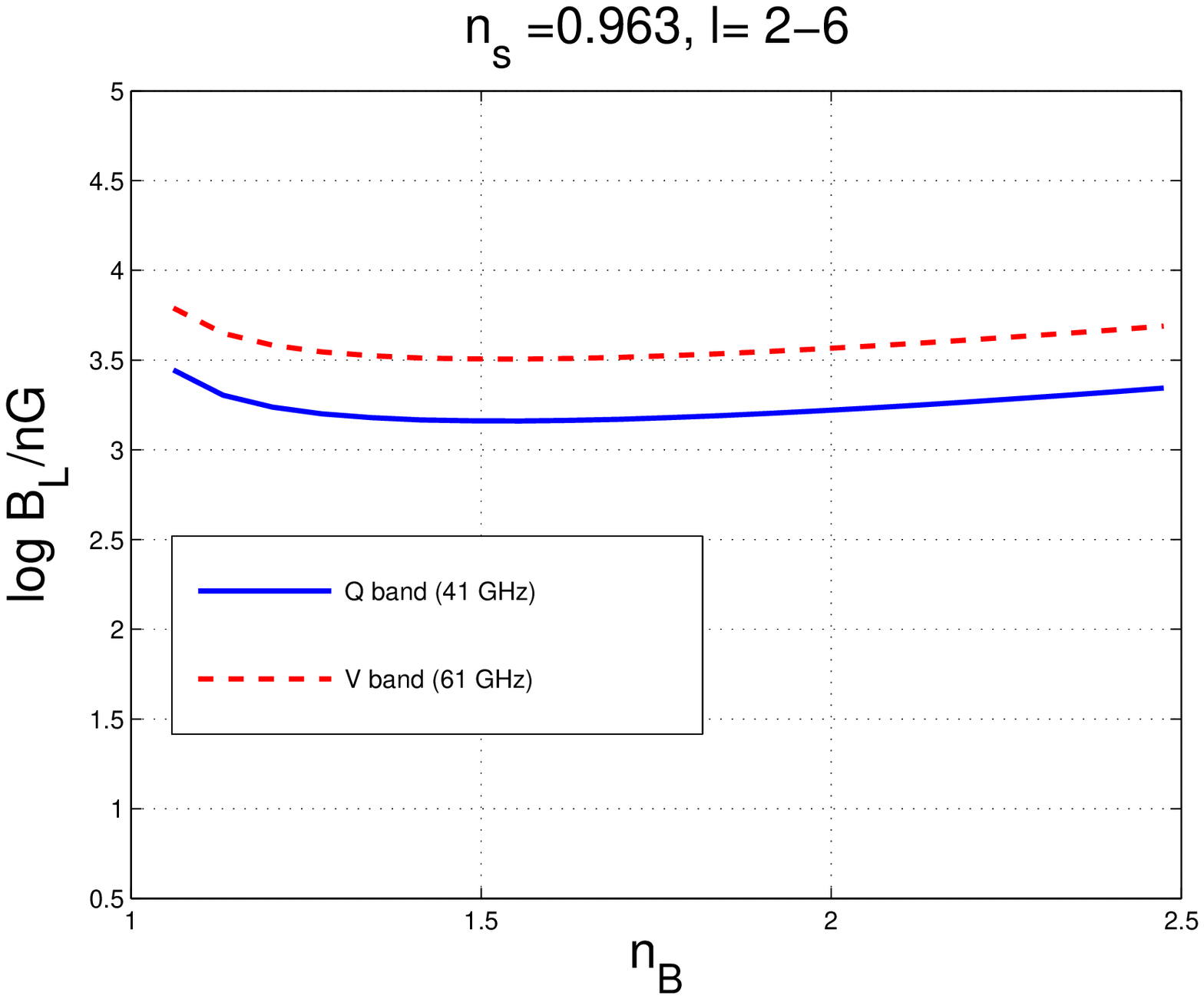}
\caption[a]{The WMAP upper limit applied to the case of the magnetized B-mode.}
\label{Figure5}      
\end{figure}
\begin{figure}
\begin{center}
\begin{tabular}{|c|c|}
      \hline
     \includegraphics[height=8cm]{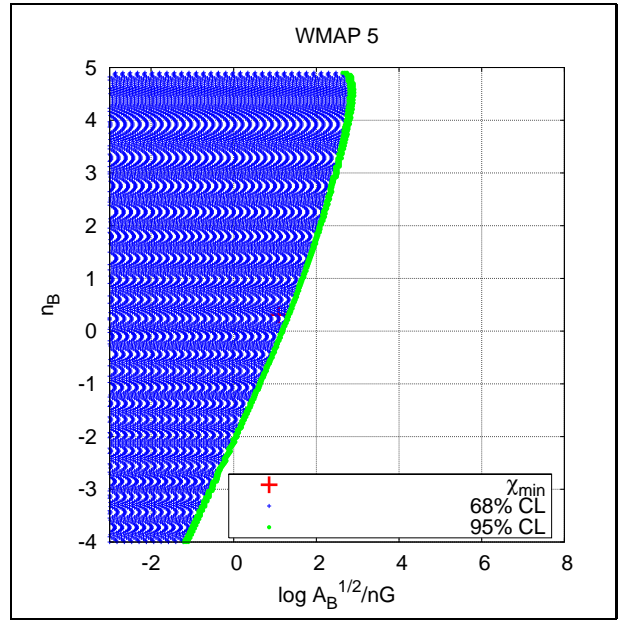}\\
      \hline
\end{tabular}
\end{center}
\caption{Confidence regions using WMAP5yr data \cite{WMAP51,WMAP52,WMAP53,WMAP54,WMAP55}.}
\label{Figure6}
\end{figure}
Since some authors insist in setting limits only for the case of blue spectral indices, in Fig. \ref{Figure4} 
the upper limits of the various experiments are expressed in the plane $(n_{\mathrm{B}},\, B_{\mathrm{L}})$ but 
only in the case of $n_{\mathrm{B}} >1$. The exactly scale-invariant case (i.e. $n_{\mathrm{B}} =1$) 
would lead to a logarithmic divergence in the two-point function in real space. This is the same 
kind of logarithmic divergence encountered in the case of curvature perturbations 
in the case of the (exact) Harrison-Zeldovich limit (i.e. $n_{\mathrm{s}} =1$). In the latter case we do have that 
\begin{equation}
\langle {\mathcal R}(\vec{x},\tau_{0}) {\mathcal R}(\vec{y},\tau_{0}) \rangle = 
\int \frac{d k}{k} {\mathcal P}_{\mathcal R}(k) \frac{\sin{kr}}{kr} \equiv \int {\mathcal A}_{\mathcal R}\frac{d k}{k}  \frac{\sin{kr}}{kr},
\label{PH7}
\end{equation}
where the second equality follows from the first by recalling that ${\mathcal P}_{{\mathcal R}} = {\mathcal A}_{{\mathcal R}} (k/k_{\mathrm{p}})^{n_{\mathrm{s}}-1}$ and by setting $n_{\mathrm{s}}=1$.  
As in the conventional $\Lambda$CDM paradigm the point $n_{\mathrm{s}}=1$ is not artificially excluded 
from the analyses, for the very same reason the point $n_{\mathrm{B}} =1$ is by no means less physical 
than the region $n_{\mathrm{B}} > 1$ and must be considered when scanning  the parameter space of the magnetized $\Lambda$CDM model. 
Also the WMAP collaboration published an upper limit on the B-mode autocorrelations and we are 
now going to compare such a bound with the magnetized B-mode.  The WMAP 
collaboration states an upper limit at $95 \%$CL which can be phrased as follows:
\begin{equation}
\frac{\ell (\ell +1) C_{\ell}^{\mathrm{BB}}}{2\pi} \leq 0.15 (\mu\, \mathrm{K})^2,\qquad 
2 \leq \ell \leq 6.
\label{PH8}
\end{equation}
 Since to derive the upper limit only the Q and V channels have been used the bounds should be imposed 
 both at $41$ and at $61$ GHz knowing that, however, the most constraining frequency will always 
 be the smallest. In Fig. \ref{Figure5} the WMAP upper limit is applied to the 
 magnetized B-mode. The resulting constraint is extremely weak and it is consistent with  published 
 bounds \cite{gk3}. 
\begin{figure}
\begin{center}
\begin{tabular}{|c|c|}
      \hline
\includegraphics[height=8cm]{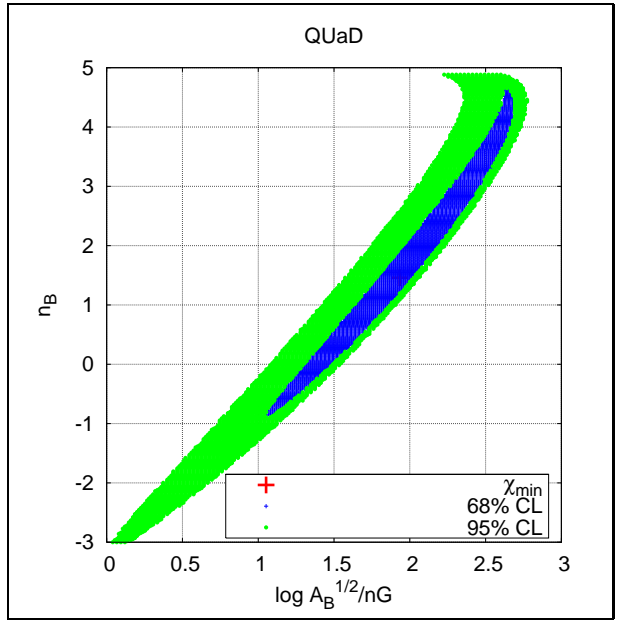}  \\
      \hline
\end{tabular}
\end{center}
\caption{Confidence regions using Quad data.}
\label{Figure7}
\end{figure}
So far the upper limits provided by the various experiments have been taken at face value and 
the consequent bounds on the parameter space of the large-scale magnetic field have 
been derived. It is also possible to use the data on the B-mode polarization 
in a dedicated $\chi^2$ analysis. The idea pursued here is rather simple and it is based 
on the observation that, nowadays, the differences between 
bayesian and Gaussian (i.e. frequentist) approaches are becoming less relevant. 
For instance, in \cite{WMAPthird1} the joint two-dimensional marginalized contours 
at $68\%$ and $95\%$ CL have been reported for various combinations of parameters 
in the case of the WMAP data alone (see, e. g. Fig. 10 in Ref. \cite{WMAPthird1}).
The various marginalized contours are ellipses which have, approximately, the 
Gaussian dependence on the confidence level. 

In this situation it is legitimate to consider all the standard $\Lambda$CDM 
parameters as Gaussian variables whose mean is given by the 
best-fit value, for instance, of the WMAP 5-yr data alone: 
\begin{equation}
(\Omega_{\mathrm{b}0},\, \Omega_{\mathrm{c}0}, \,\Omega_{\mathrm{\Lambda}}, \,h_{0}, \,n_{\mathrm{s}},\,\tau)= (0.0441,\, 0.214,\, 0.742,\, 0.719,\, 0.963,\,0.087).
\label{best1}
\end{equation}
Following this frequentist approach the parameters of the magnetized background can be 
estimated by minimizing 
\begin{eqnarray}
\chi^2(n_\mathrm{B},A_\mathrm{B})=\sum_{i=1}^{N_{data}}\left(\frac{C_{\ell}^{(\mathrm{BB})}(i)-
C_{\ell}^{(\mathrm{BB})}(i,n_\mathrm{B},A_\mathrm{B})}{\sigma(i)}\right)^2,
\end{eqnarray}
where the sum goes over all data points; $C_{\ell}^{(\mathrm{BB})}(i)$ and $\sigma(i)$ are  the 
observational data and their binned errors. 
Finally $C_{\ell}^{(\mathrm{BB})}(i,n_\mathrm{B},A_\mathrm{B})$ are the values of the magnetized B-modes computed
using the approach outlined in the previous sections. 
\begin{figure}
\begin{center}
\begin{tabular}{|c|c|}
      \hline
  \includegraphics[height=8cm]{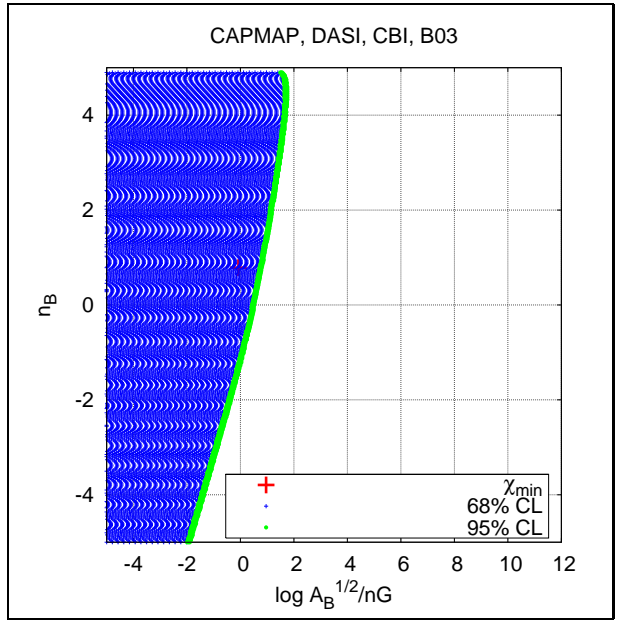} \\
      \hline
\end{tabular}
\end{center}
\caption{Confidence regions using  the combined data sets of Capmap, Dasi, Cbi  
and B03.}
\label{Figure8}
\end{figure}
The $(n_{\mathrm{B}}, A_{\mathrm{B}})$ - parameter space is explored using a grid of $450^2$ points. 
In the following the contour plots for 68\% and 95\% confidence level are presented for different data sets. Since the instruments operate at different frequency channels, however, the data are usually those for combined frequency channels,  it seems useful to assign effective frequencies to each of the experiments. Following 
\cite{erik} an effective frequency $\nu_{\mathrm{eff}}$ can be defined by
\begin{eqnarray}
S(\nu_{\mathrm{eff}})=\int f(\nu)S(\nu)d\nu,
\end{eqnarray}
where $S(\nu)$ is the frequency spectrum of the signal, and $f(\nu)$ is the frequency response profile of the detector. The polarization signal is determined by $\Delta_{\mathrm{B}}(\hat{n})$ which is proportional to $\nu^{-2}$, so that the spectrum of the signal is given by
\begin{eqnarray}
S(\nu)=S_0\left(\frac{\nu}{\nu_{\mathrm{max}}}\right)^{-2}.
\end{eqnarray}
There are different options for the detector response function. For a flat bandpass filter with sharp
cut-offs at some frequencies $\nu_a$ and $\nu_b$ the effective frequency is given by
$\nu_{\mathrm{eff}}=\sqrt{\nu_a\nu_b}$.  Applying this to the Dasi and Cbi experiments gives in both cases
$\nu_{\mathrm{eff}}=31$ GHz. Assuming that the detector response function is modelled by a $\delta$ function at the relevant frequencies it is found that the effective frequency is given by $\nu_{\mathrm{eff}}=\frac{\sqrt{2}\nu_a\nu_b}{\sqrt{\nu_a^2+\nu_b^2}}$, which in the case of QUaD results in $\nu_{eff}=118$ GHz and the Q+V data of WMAP in $\nu_{eff}=48$ GHz. For Capmap combining the two detector response functions an effective frequency of $\nu=52$ GHz is found.

We are now ready to discuss the main results of this analysis.
In Fig. \ref{Figure6} results are presented for WMAP 5. The minimum value of $\chi^2$ corresponds to $A_B^{\frac{1}{2}}=12.89$ nG and $n_B=0.35$ which corresponds to a magnetic field strength $B_L=332$ nG. Furthermore, $\chi^2_{\mathrm{min}}=5.8$ and there are $8$ data points. 
In Fig. \ref{Figure7} the results are presented for the Quad data which consist of 23 data points. It is found that the bestfit model has $\chi_{\mathrm{min}}^2=11.65$ and corresponds to the magnetic field parameters $A_B^{\frac{1}{2}}=88$ nG, $n_B=1.49$. This implies a magnetic field strength
$B_L=108$ nG.
By comparing Fig. \ref{Figure7} and \ref{Figure6} 
it seems that the Quad data lead to systematically larger values of 
the magnetic fields. At the same time, by admission of the 
Quad team, the BB angular power spectra might not be free 
of systematics. Indeed, the argument used by the Quad team seems to be, in short, the following (see \cite{quad2}). If Quad would have seen BB power at a level greater than in the $\Lambda$CDM, extensive investigations 
would have been required to check if the result was indeed free 
of systematics. Since the collaboration did not see any excess in the BB power, ``further investigation of mixings systematics is arguably 
unnecessary" (verbatim from \cite{quad2}). The caveat is that, in the present case, the putative signal is larger than in the $\Lambda$CDM case and, probably, further checks on the systematics may resolve 
the problem.

The results for the combined data sets of Capmap, Dasi, Cbi and Boomerang B03 are presented in Fig. \ref{Figure8}. In total there are 24 data points. The bestfit model is characterized by $\chi_{\mathrm{min}}^2=12.93$ and the magnetic field parameters, 
$A_B^{\frac{1}{2}}=0.8$ nG and $n_B=0.76$. This results in a magnetic field strength $B_L=6.23$ nG.
Concerning the results expressed by Figs. \ref{Figure6}, \ref{Figure7} and \ref{Figure8} few comments are in order.  
In all cases the relevant statistical indicator, for the present analysis, is the $\Delta \chi^2$, i.e. 
$\chi^2 - \chi^2_{\mathrm{min}}$. We do not judge here the compatibility of a model with the data 
by performing the Pearson $\chi^2$ test. 

\section{Concluding remarks}

In summary, the B-mode due to Faraday rotation has been calculated both semi-analytically and numerically. The parameter space of the magnetized background has been investigated by enforcing the upper limits on the B-mode polarization and also by performing a dedicated $\chi^2$ analysis. The two 
analyses led to compatible results. 
It has been shown that forthcoming experiments aiming at a direct detection of a B-mode polarization arising from a putative tensor power spectrum can also be able 
to cut through the parameter space of a magnetized 
background. 

An interesting byproduct of the present analysis 
has been an analytic scrutiny of the Faraday induced B-mode at small angular scales.  In the small-scale limit 
the B-mode autocorrelations can be computed with reasonable accuracy by contracting 
(in the In\"on\"u-Wigner sense) the relevant matrix  
elements of the rotation group. The latter results have been cross-checked by direct numerical calculations.  
\section*{Acknowledgments}
It is a pleasure to acknowledge 
useful discussions with L. Alvarez-Gaum\'e. 
K.E.K. is supported by the ``Ram\'on y Cajal''  program and by the grants FPA2005-04823, FIS2006-05319 and CSD2007-00042 of the Spanish Science Ministry.
M. G. wishes to thank T. Basaglia of the CERN scientific 
information service for providing copies of 
Refs. \cite{vilenkin,ponzano}.

\end{document}